\documentclass{article}
\usepackage[T1]{fontenc}
\usepackage{preprint,times}

\usepackage{amsmath,amsfonts,bm}

\def\eqref#1{equation~\ref{#1}}

\def\1{\bm{1}}

\DeclareMathAlphabet{\mathsfit}{\encodingdefault}{\sfdefault}{m}{sl}
\SetMathAlphabet{\mathsfit}{bold}{\encodingdefault}{\sfdefault}{bx}{n}

\usepackage{amsmath}
\usepackage{amssymb}
\usepackage{booktabs}
\usepackage{graphicx}
\usepackage{xcolor}
\usepackage{hyperref}
\hypersetup{hidelinks,pdftitle={SPLASH: Switching Parallel Layouts of Attention with Seamless Handoff for LLM Serving},pdfauthor={Chuan Liu, Shuoming Zhang, Zhicheng Li, Qianqi Sun, Ruiyuan Xu, Qiuchu Yu, Xiyu Shi, Huimin Cui, Jiacheng Zhao}}

\usepackage{xurl}
\usepackage{array}
\usepackage{adjustbox}

\title{SPLASH: Switching Parallel Layouts\\
of Attention with Seamless Handoff\\
for LLM Serving}

\author{%
\begin{minipage}[t]{\dimexpr\textwidth-2\tabcolsep\relax}
\centering
\href{https://openreview.net/profile?id=~Chuan_Liu14}{Chuan Liu}\textsuperscript{1,2} \quad \href{https://openreview.net/profile?id=~Shuoming_Zhang1}{Shuoming Zhang}\textsuperscript{1,2,*} \quad \href{https://openreview.net/profile?id=~Zhicheng_Li5}{Zhicheng Li}\textsuperscript{1,2} \\
\href{https://openreview.net/profile?id=~Sun_Qianqi1}{Qianqi Sun}\textsuperscript{1,2} \quad \href{https://openreview.net/profile?id=~Ruiyuan_Xu1}{Ruiyuan Xu}\textsuperscript{1,2} \quad \href{https://openreview.net/profile?id=~Qiuchu_Yu1}{Qiuchu Yu}\textsuperscript{1,2} \\
\href{https://openreview.net/profile?id=~Xiyu_Shi2}{Xiyu Shi}\textsuperscript{1} \quad \href{https://openreview.net/profile?id=~Huimin_Cui1}{Huimin Cui}\textsuperscript{1,2} \quad \href{https://openreview.net/profile?id=~Jiacheng_Zhao1}{Jiacheng Zhao}\textsuperscript{1,2} \\[0.5em]
{\normalfont\small
\textsuperscript{1}State Key Laboratory of Processors, Institute of Computing Technology\\
Chinese Academy of Sciences, Beijing, China\\
\textsuperscript{2}University of Chinese Academy of Sciences, Beijing, China}
\end{minipage}%
}

\iclrfinalcopy
\begin{document}
\raggedbottom

\maketitle
\lhead{}
\renewcommand{\headrulewidth}{0pt}
\begingroup
\renewcommand{\thefootnote}{}
\footnotetext{Preprint. Under review.}
\renewcommand{\thefootnote}{*}
\footnotetext{Corresponding to: \texttt{zhangshuoming21s@ict.ac.cn}.}
\endgroup
\begin{abstract}
No single way of parallelizing attention serves large language models well under all loads. Low concurrency favors tensor parallelism, many independent requests favor data-parallel attention, and long prompts favor context parallelism. Reasoning, agentic, and RL-rollout workloads make a fixed choice untenable: a batch that begins as many short requests ends as a few very long ones, 
so the best layout changes while the same requests run. Serving engines nevertheless fix one layout at launch, because changing it has meant draining requests and restarting workers. We present \textsc{Splash}, a serving system that switches the parallel layout of attention while requests are running. It builds on one observation: modern attention, with few or no KV heads, decouples where a request's KV cache lives from how attention weights are sharded. This has two consequences. First, layouts differ only in who owns the weights and the cache, and most of that state already sits where the next layout needs it; \textsc{Splash} reuses it, moves the rest in the background of ongoing inference, and hands off at a batch boundary, 
making a switch nearly free: its median overhead is under 0.51\% of the step it runs in. Second, the decoupling exposes a layout that existing engines lack: Decoupled Ownership Parallelism (DOP) shards attention weights as tensor parallelism does while keeping each request's cache on a single owner as data-parallel attention does. DOP replicates neither, 
offers 27--60\% more KV capacity than data-parallel attention, and gives the scheduler a choice when KV memory limits admission. A transition-aware scheduler follows the best of the four layouts as load changes. 
On B200 GPUs serving GLM-5.3, \textsc{Splash} improves end-to-end serving throughput by 1.3--1.73$\times$ over fixed-layout deployments, and the same layout regimes appear with DeepSeek-V3.2 on H200 and GLM-5.3-Flash on DCU.
Our implementation is open source at
\url{https://github.com/ict-agent/SPLASH-sglang}.
\end{abstract}

\section{Introduction}
\label{sec:introduction}

Agents have turned language models from something people ask a question into something they run all day \citep{agentix,agentworkload}. A serving system must turn a fixed set of GPUs into as many answered requests as it can, whatever requests arrive.
The load keeps moving: both the number of requests in flight and the length of their histories change while the system runs. The sharpest case is reinforcement-learning rollout \citep{deepseekr1,kimik15}: each step launches thousands of generations at once, most finish quickly, and a few long reasoning traces hold the whole group until they end \citep{seer,rollpacker}.

\begin{figure}[t]
\centering
\includegraphics[width=\linewidth]{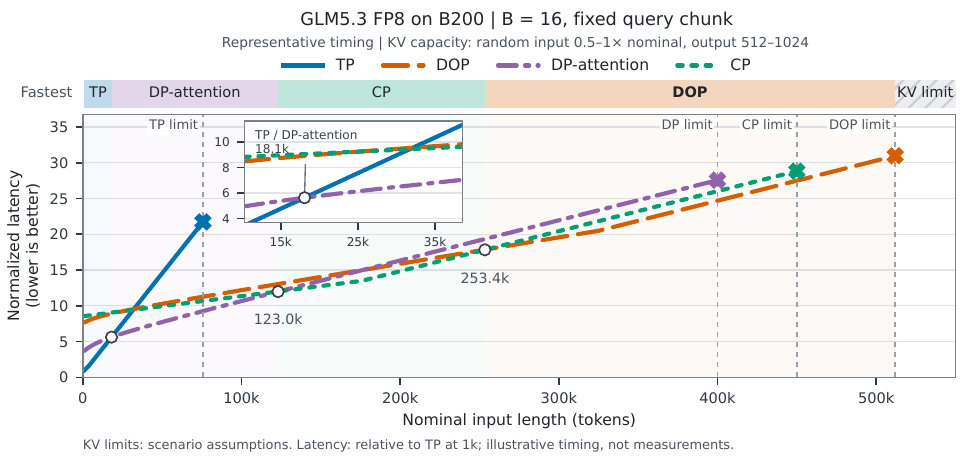}
\caption{Conditional latency model for GLM-5.3 FP8 on B200 GPUs
($B=16$, fixed query chunk), normalized to TP at 1k. Open circles mark
changes in the fastest feasible layout; the inset enlarges the first.
Crosses mark KV capacity limits, and hatched regions indicate where the KV cache exceeds the available capacity. \autoref{app:inference-model}
gives the timing and capacity assumptions.}
\label{fig:inference-bottleneck-latency}
\end{figure}

Attention is where that variation forces a choice among three layouts:
tensor parallelism (TP) \citep{megatronlm2019} for few requests in flight,
context parallelism (CP) \citep{cp-million} for a prompt that outgrows one
rank, and data-parallel attention (DP-attention)
\citep{sglangv04,deepseekinference} for many independent requests
(\autoref{fig:layout-tradeoffs}). Each trades compute, communication, and
cache room differently, so each wins under some load and loses under another
(\autoref{fig:inference-bottleneck-latency}); a choice made at launch cannot
follow a load that keeps moving.

What stands in the way is state that cannot be rebuilt cheaply. Weights are read-only: a rank can view them differently, or receive them once, and be done. 
Live KV histories, by contrast, keep growing while the engine serves.
Changing who owns a cache has to preserve every history, the order things were appended in, and the sequence of collectives that all ranks agree on. 
Engines therefore drain the batch and restart the workers, which satisfies all three but takes longest exactly when a switch pays off: the long requests that make it worthwhile are the slowest to drain.

Prior systems reconfigure parallelism online \citep{loongserve,seesaw,flyingserving,shiftpar,amoeba,moebius,pat}, but they either hold the batch while the change happens or cover one pair of strategies, and a switch is not free in any of them.
Shift Parallelism \citep{shiftpar} is limited to tensor and sequence parallelism, whose caches share one head-sharded layout; MLA keeps one latent shared by all heads, so on DeepSeek-style models both layouts hold a whole copy of it on every rank and switching between them frees no cache. 
We present \textsc{Splash}, a serving system that makes \emph{attention layout} a live choice inside one worker group, covering every switch among the layouts worth running and hiding the preparation behind ongoing inference.

Sharding the attention weights used to decide where the cache lived as well. TP shards along heads, and in multi-head attention the KV cache is indexed by heads too. Multi-query and grouped-query attention cut the number of KV heads to a handful, and MLA removed them, keeping one head-shared latent history per token \citep{mqa,gqa,deepseekv2,deepseekv3}.
\textbf{Where the attention weights are sharded and which rank owns a request's KV history are now two independent choices}, 
and a layout is one way of setting both over the same request state.

Two things follow. The first is, to our knowledge, a new layout. Once placement and ownership are separate decisions, the layouts engines run are three of their combinations, and each one keeps a redundant copy of something: TP shards the projections but, with few or no KV heads left to shard, leaves a whole history on every rank; DP-attention and CP divide the history, by request and by position respectively, and leave a whole copy of the projections. 
Sharding the projections, keeping each request's history on one owner, and
exchanging activations between them on every attention layer removes both
copies; we call the result \textbf{Decoupled Ownership Parallelism (DOP)}. It
holds the largest cache of the four on a given group, the room decode needs
once long contexts fill memory, as on accelerators with 64--192\,GB per
device.

The second is that a switch between any two of the four can be planned as a difference: what a rank already holds, set against what the destination needs. 
Moving from DP-attention to DOP, a rank keeps its KV and cuts DOP's shards out of the full projections it already holds; moving from TP to DOP, it keeps one owner's copy of each MLA history and retires the rest. Neither sends a byte over the network.
The cost of a switch thus follows the state the destination lacks, not the size of the model or the cache. 
One account covers all $A_4^2=12$ directed switches, one per ordered pair of layouts, and three collectives supply every one of them: discard, all-gather, and all-to-all. \textsc{Splash} moves the missing state one layer at a time inside an ordinary serving step: layer $i$ receives its new weights and KV while layer $i-1$ computes. A transition-aware scheduler decides which layout to run and when a switch is worth making. In a rollout, the group runs DP-attention while the batch is wide and CP once only the long tail remains.

\paragraph{Contributions.}
\begin{itemize}
  \item \textbf{An ownership model} that reduces TP, CP, DP-attention, and DOP to two choices, where the projections live and who owns the cache, and derives from them each layout's per-rank footprint, its longest feasible context on a fixed group, 
the state each of the $A_4^2$ directed switches retains, and the collective that supplies the rest: discard, all-gather, or all-to-all.
  \item \textbf{Decoupled Ownership Parallelism}, the zero-redundancy layout that the ownership model exposes: it stores every attention weight and every history once, pays a per-layer activation exchange that stays constant as contexts grow, and frees 12.69\,GiB per GPU for KV over DP-attention on 
GLM-5.3 (B200 GPUs, FP8).
  \item \textbf{\textsc{Splash}}, a serving system that moves a running batch among all four layouts 
inside an ordinary serving step. On B200, one layer's communication takes 0.29--0.30\,ms for weights and 3.47--3.81\,ms for KV at $B=16$ and 256K tokens, while a complete switch adds 0.02--11.76\,ms end to end, against up to 667.31\,ms when it blocks, 
because later layers' transfers hide behind computation. Its transition-aware scheduler runs the fastest feasible layout and switches once another leads by a set margin, so the deployment follows the load.
\end{itemize}

\section{An Ownership Model of Attention Layouts}
\label{sec:motivation}
\label{sec:layouts}

This section makes the two-choice view of \autoref{sec:introduction} precise
and derives from it each layout's per-rank memory (\autoref{sec:dop}) and the
state a switch must supply (\autoref{sec:method}).

\begin{figure}[t]
\centering
\includegraphics[width=\linewidth]{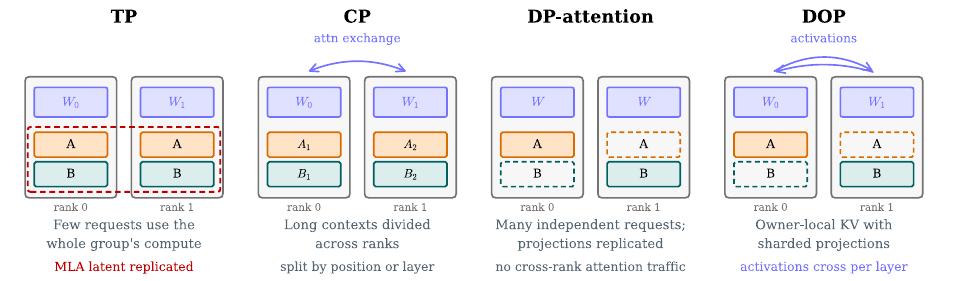}
\caption{Four attention layouts on two ranks serving requests A and B.
Blue blocks are projections, orange and teal bars the KV of A and B,
dashed blocks are absent from a rank, 
the red outline marks MLA's history replicated under TP, 
and arrows mark attention's cross-rank traffic.}
\label{fig:layout-tradeoffs}
\end{figure}

We consider $T$ ranks serving one model whose dense and expert layers keep
their parallelism while attention changes layout. A layout decides where
the large attention projections live, sharded across the ranks or replicated
on each, 
and which ranks hold each request's KV history
(\autoref{fig:layout-tradeoffs}). Under MLA the two choices are independent:
the latent cache has no head axis, so TP's head-sharded projections leave a full copy of every
history on every rank (red outline).

\paragraph{Footprint.}
Let $W_A$ be the size of the projections that TP and DOP shard, $k$
the KV bytes per context token over all cache-bearing layers
($L(d_c+d_r)b$ for absorbed MLA with latent width $d_c$, rotary width $d_r$,
and $b$ bytes per element), and $B$ the number of balanced requests with
context $s$. Each rank holds
\begin{equation}
\begin{aligned}
 M_{\mathrm{TP}}(s)&=W_A/T+Bks, &
 M_{\mathrm{DOP}}(s)&=W_A/T+Bks/T,\\
 M_{\mathrm{DP}}(s)&=M_{\mathrm{CP}}(s)=W_A+Bks/T.
\end{aligned}
\label{eq:inference-layout-memory}
\end{equation}
Every existing layout stores one redundant copy: TP of each history,
DP-attention and CP of the projections. DOP stores each projection and history
once, so under a per-rank budget
$M_{\mathrm{avail}}$, it reaches the longest balanced context,
$(TM_{\mathrm{avail}}-W_A)/(Bk)$, against $T(M_{\mathrm{avail}}-W_A)/(Bk)$
for DP-attention and CP and $(M_{\mathrm{avail}}-W_A/T)/(Bk)$ for TP.
Execution cost, by contrast, orders the layouts differently and shifts with
context length
(\autoref{fig:inference-bottleneck-latency};
\autoref{app:inference-model}).

\paragraph{Switches.}
\label{sec:reuse}
All four layouts hold the same weights and histories, so a switch between
two panels of \autoref{fig:layout-tradeoffs} reduces to changing which blocks
each rank holds. Let $\mathcal K_r^a$ and $\mathcal P_{r,k}^a$ be
the KV blocks and layer-$k$ projection blocks that rank $r$ holds in layout
$a$. A switch to $b$ 
keeps every block the two layouts share and supplies the difference
\begin{align}
 V_r^K(a\!\to\!b)
 &=\big|\mathcal K_r^b\setminus\mathcal K_r^a\big|_{\mathrm{bytes}},
 \label{eq:kv-missing}\\
 V_{r,k}^W(a\!\to\!b)
 &=\big|\mathcal P_{r,k}^b\setminus\mathcal P_{r,k}^a\big|_{\mathrm{bytes}}
 \label{eq:weight-missing}
\end{align}
to rank $r$. Three collectives supply every such difference
(\autoref{fig:splash}, right). A rank \emph{discards} when the destination
needs a subset of its blocks: 
it keeps those and frees the rest in place, at zero network cost. It \emph{all-gathers} blocks spread over the other ranks, and ranks
exchange blocks \emph{all-to-all} when state moves between request ownership
and position sharding. Weights all-gather $(1-1/T)$ of each layer's
projections per rank when TP or DOP switches into replicated projections, a
volume the model fixes, and are discarded in every other switch. KV
all-gathers into TP, moves all-to-all between CP and the request-owned
layouts, and is discarded in the remaining five switches; 
its volume grows linearly with the live context $Bs$.

\section{DOP: Sharded Projections with Request-Owned KV}
\label{sec:dop}

\autoref{sec:layouts} 
leaves one combination with no redundant copy: projections sharded as in TP, and 
each request's history 
on one owner as in DP-attention, chosen at admission from projected KV growth. 

\paragraph{Computation.}
Sharded projections give each rank all $N$ current query rows for its $H/T$
heads, $[N,H/T,d_q]$, whereas attention at an owner needs its own
$N_r$ rows for all $H$ heads, $[N_r,H,d_q]$. DOP converts between the two
with a variable-size all-to-all before attention and a reverse one after it,
which returns head slices to the sharded value and output projections 
(arrows in \autoref{fig:layout-tradeoffs}). Every head still attends over
the complete history, so DOP's output matches TP's up to floating-point
reduction order. 
Unlike DeepSpeed-Ulysses \citep{ulysses}, whose all-to-all regroups a sequence by head and therefore needs a cache split by head, DOP regroups rows by request and keeps each head-free MLA history whole on one owner.

\paragraph{Cost.}
With returned per-head width $d_z$, the two
exchanges of one layer receive
\begin{equation}
 V_{\mathrm{DOP}}=(1-1/T)NH(d_q+d_z)b
 \label{eq:dop-traffic}
\end{equation}
bytes in aggregate, with $d_q=d_c+d_r$ and $d_z=d_c$ for absorbed MLA. 
This volume tracks the step's query rows, one per request in decode and the chunk in prefill, independent of context length. 

\paragraph{Capacity.}
In return, DOP leaves $W_A(1-1/T)$ more bytes per rank for KV than DP-attention (\autoref{eq:inference-layout-memory}). 
For GLM-5.3 in FP8 at $T=8$, 
sharding the Q-B, KV-B, and O projections of all 78 layers 
frees 12.69\,GiB per rank for KV (\autoref{tab:capacity}), memory DP-attention spends on replicas.

\paragraph{When DOP pays off.}
The returned memory raises throughput once it changes admission
(\autoref{sec:capacity-eval}), and the exchange of \autoref{eq:dop-traffic}
is its price, so DOP leads where contexts are long and batches are large
while DP-attention leads over the middle of both axes
(\autoref{sec:regimes}). DOP therefore complements DP-attention, and the
two are cheap to switch between: 
DOP shares projection shards with TP and request owners with DP-attention, so a switch into DOP from either one is a pair of discards 
(\autoref{fig:splash}, right). A variant, \emph{DOP-P}, caches some full projections on the owner, trading part of the returned memory for less exchange. The same
design applies to MQA and GQA, where its gain depends on how much of the
cache TP already splits.

\section{\textsc{Splash}: Live Switching Among Four Layouts}
\label{sec:method}
\label{sec:hot-switch}

\begin{figure}[t]
\centering
\includegraphics[width=\linewidth]{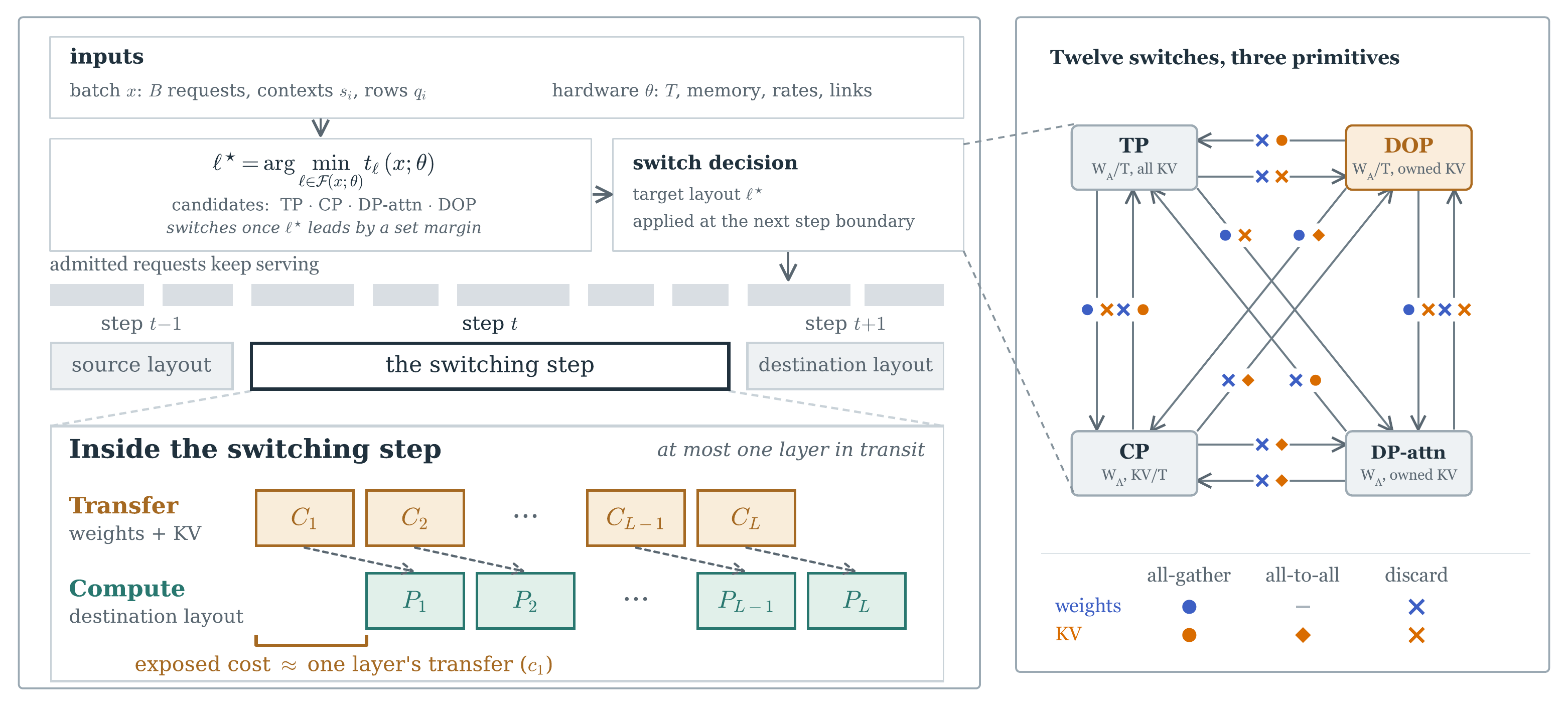}
\caption{\textsc{Splash}. Left: the scheduler runs the fastest feasible layout
and switches once another leads by a set margin (\autoref{sec:policy}).
Admitted requests keep serving: layer $i$'s transfer ($C_i$) overlaps
layer $i-1$'s computation ($P_{i-1}$), exposing about one layer's transfer
($c_1$) with at most one layer in transit.
Right: the $A_4^2=12$ directed switches; 
arrows show the primitives for weights (blue) and KV (orange). A discard
keeps what the destination needs and frees the rest at zero network cost.}
\label{fig:splash}
\label{fig:transitions}
\end{figure}

\autoref{sec:layouts} fixes what each switch must supply; \textsc{Splash}
supplies it while serving continues, moving the missing state one layer ahead
of the computation (\autoref{fig:splash}), and \autoref{sec:policy} decides
when to switch.

\subsection{Switching within one step}
\label{sec:state-interface}
\label{sec:prepare}

\paragraph{1. Reuse.}
At the step boundary, each rank 
evaluates \autoref{eq:kv-missing} and \autoref{eq:weight-missing} 
into a manifest assigning each layer's weights and KV one primitive
(\autoref{fig:splash}, right): a discard keeps the blocks the destination
needs, 
repacked into its view, and frees the rest, while an all-gather or all-to-all fetches the missing blocks. Reuse requires matching precision, quantization scales, and position convention. 

\paragraph{2. Overlapped transfer.}
Layer~1's missing state transfers first; 
then layer $i$'s collectives run on a separate stream while layer $i-1$ computes. When a switch enters replicated projections, 
the weights in flight are one layer's projections: by the analytical count of \autoref{app:weights}, 145.73\,MiB per rank for GLM-5.3 in FP8 at $T=8$, against 11.3\,GiB for all 78 layers. Because the switch starts at a step boundary and each layer moves before it computes, every layer arrives as a complete, static snapshot.

\paragraph{3. Handoff.}
\label{sec:handoff}
Layer $i$ runs in the destination layout once its state arrives. As all ranks switch in one step, every layer's collectives match across ranks. Requests keep their histories and workers keep running; every query reads its complete causal prefix, and indexer, position, and recurrent
state move with the layer's KV. 
Later steps replay the destination's graphs, captured at startup.

\paragraph{4. Retire.}
A layer's old pages and projection replicas are freed once its transfer completes. With at most one layer in transit and $\Delta_r^{\mathrm{layer}}$ its buffers, rank $r$ therefore stays within
\begin{equation}
 \max\{M_r(a),M_r(b)\}+\Delta_r^{\mathrm{layer}}+R_r\leq H_r,
 \label{eq:transition-memory}
\end{equation}
where $M_r(a)$ and $M_r(b)$ are the source and destination footprints, $H_r$
is device memory, and $R_r$ a reserve. \autoref{app:limitations} details
the remaining costs.

\subsection{Transition-aware scheduling}
\label{sec:policy}
\label{sec:switch-cost}

The scheduler is a function $\pi(x;\theta)$. The hardware constants $\theta$ are
the group size $T$, device memory $H_r$, compute rate, 
weight-read, KV-read, and interconnect bandwidths, and fixed collective latency, calibrated once
per deployment from microbenchmarks. The current batch $x$ is the number of
running requests $B$, each request's live context $s_i$, and its query rows $q_i$
in this step: the chunk length in prefill and one in decode. The output is one of
the four layouts with its projection caching, CP storage, backend, and chunk size.

A candidate layout $\ell$ must fit in steady state,
\begin{equation}
 M_{0,r}+W_r(\ell)+K_r(\ell,x)+A_r(\ell,x)+R_r\leq H_r,
 \label{eq:feasible}
\end{equation}
where $M_0$ is common state, $W$ the projection footprint, $K$ the live KV
projected from the current batch, and $A$ scratch and graphs; the switch from
the current layout must also satisfy \autoref{eq:transition-memory}. 
The two constraints define the feasible set $\mathcal F(x;\theta)$.

For each feasible $\ell$, the ownership model fixes its load on every rank:
the projection footprint $W$ is the whole weights or a $1/T$ shard, and KV,
attention work, and KV reads each count $B$, $B/T$, or $\lceil B/T\rceil$
requests (\autoref{tab:rank-loads}). The stage-wise cost model of
\autoref{app:inference-model} (\autoref{eq:inference-bottleneck-cost}) turns
these loads and $\theta$ into the per-step latency $t_\ell(x;\theta)$;
\autoref{fig:inference-bottleneck-latency} is its slice at $B=16$ over context
length.

A switch exposes the first layer's transfer and any later transfer that
outlasts the previous layer's computation:
\begin{equation}
 \widehat t_{\mathrm{exposed}}=\widehat t_{\mathrm{xfer},1}
 +\sum_{i=2}^{L}\big[\widehat t_{\mathrm{xfer},i}
 -\widehat t_{\mathrm{comp},i-1}\big]_+ .
 \label{eq:switch-cost}
\end{equation}
One layer's communication costs 0.29--0.30\,ms for weights and
3.47--3.81\,ms for KV at $B=16$ and 256K on B200
(\autoref{sec:switch-microbench}); because later layers hide behind
computation, a complete switch adds 0.02--11.76\,ms end to end, under 0.51\%
of the step it runs in at the median (\autoref{sec:live-overhead}). Per-step
latency alone therefore ranks the layouts, and the switch cost enters as a
margin.

\textsc{Splash} picks
\begin{equation}
 \ell^\star=\arg\min_{\ell\in\mathcal F(x;\theta)} t_\ell(x;\theta),
 \label{eq:policy}
\end{equation}
and re-evaluates the choice as the batch evolves, replacing the current layout once $\ell^\star$ leads it by a set margin; the margin keeps the layout stable while the load hovers near a crossover.

Batch size and admission are inputs to $\pi$; choosing them jointly with the
layout is future work.

\section{Evaluation}
\label{sec:evaluation}
\label{sec:implementation}

\subsection{Experimental setup}
\label{sec:scope}

\paragraph{Models and software.}
\textsc{Splash} is built atop SGLang 0.5.10. We compare TP, CP,
DP-attention, and DOP using GLM-5.3 on B200, GLM-5.3-Flash on DCU, and
DeepSeek-V3.2 on H200; the main text analyzes GLM-5.3 with 78 layers, FP8
weights, and FP8 KV.
The DCU and H200 experiments appear in \autoref{app:dcu-evaluation}
and \autoref{app:h200-evaluation}, respectively.

\paragraph{Hardware and deployment.}
Each B200 deployment separates prefill and decode: eight GPUs on one
node run prefill, and sixteen GPUs across two nodes run DP-attention
decode. Prefill layouts share the same decode configuration. \autoref{app:repro} gives detailed
deployment settings. The B200 switch studies use $B=16$ and 256K context
tokens per request.

\subsection{Fixed-layout performance}
\label{sec:regimes}

Four B200 sweeps cover eight input lengths from 1K to 512K at
$B=16,32$, and eight batch sizes
$B\in\{1,4,8,16,32,64,128,256\}$ at 64K and 128K input.
All four layouts run at every position, giving 128 measurements.
$B$ is both the request count and client concurrency; every request
generates 1,024 tokens, and K denotes 1,024 tokens. Total throughput is
$B(L_{\mathrm{in}}+1{,}024)/T_{\mathrm{batch}}$, including queueing,
prefill, KV transfer, and decode. 
Each point is the mean of 10 runs (\autoref{app:full-sweep}).

\begin{figure}[t]
\centering
\includegraphics[width=\linewidth]{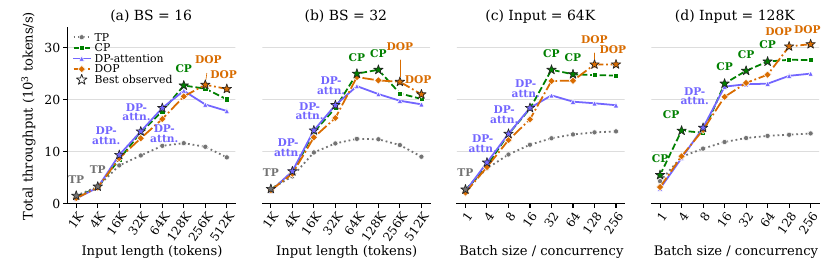}
\caption{Fixed-layout GLM-5.3 performance on B200 with FP8 weights,
common DP-attention decode, and 1K output tokens per request.
(a, b) Input-length sweeps at $B=16,32$; (c, d) batch-size sweeps at
64K and 128K input. Sampled positions are equally spaced.
Stars and labels mark the best recorded layout at each position.}
\label{fig:regime-crossover}
\label{fig:regime-batch}
\end{figure}

\paragraph{Input length selects the layout.}
At $B=16$, TP leads at 1K--4K, DP-attention at 16K--64K,
CP at 128K, and DOP at 256K--512K (\autoref{fig:regime-crossover}a), the
order the cost model of \autoref{fig:inference-bottleneck-latency} predicts;
its last two crossovers, at 123.0k and 253.4k tokens, fall between the
measured positions.
At $B=32$, TP leads at 1K, DP-attention at 4K--32K, CP at
64K--128K, and DOP at 256K--512K (\autoref{fig:regime-crossover}b).
Across the four longest-input points, DOP reaches 21,059--23,470
tokens/s, 3.4--10.6\% above CP, the runner-up.

\paragraph{Concurrency reorders the ranking.}
At 64K input, TP leads at $B=1$, DP-attention at $B=4,8,16$,
CP at $B=32,64$, and DOP at $B=128,256$ (\autoref{fig:regime-crossover}c),
so the batch sizes a rollout passes through as it drains from 256 requests to
one span all four regimes.
At 128K, CP leads at $B=1,4,16,32,64$, DP-attention at $B=8$,
and DOP at $B=128,256$ (\autoref{fig:regime-crossover}d).
Thus DOP first leads at $B=128$ in both batch sweeps.
Its advantage over the strongest alternative is 8.3--8.7\% at 64K
and 9.3--11.0\% at 128K; at 128K/$B=256$, it delivers 30,724
tokens/s versus CP's 27,677.

\paragraph{DOP complements the other layouts.}
DOP leads at eight of the 32 sampled positions, while TP,
DP-attention, and CP remain faster elsewhere. For example, at 128K
and $B=1,4$, DOP trails CP by 42.2\% and 35.1\%.
At $B=32$, DOP's throughput falls 3.7\% from 64K to 256K, to 23,470
tokens/s, while CP's falls 15.3\%, so DOP takes the lead.
These fixed-layout comparisons motivate workload-aware selection.

\subsection{KV capacity and request admission}
\label{sec:capacity-eval}

On B200 GPUs at memory fraction 0.85, DOP holds 8,271,360 distinct
KV tokens, versus 6,497,792 for DP-attention and 6,738,944 for CP:
27.3\% and 22.7\% more, respectively (\autoref{tab:capacity}).
TP's replicated pools hold 1,045,312 distinct tokens.
On H200 at memory fraction 0.90, DOP holds 59.7\% more distinct tokens than DP-attention%
.
On DCU with GLM-5.3-Flash, DOP retains the 19.0\% larger
per-owner KV pool (1,118,656 versus 940,224 tokens;
\autoref{tab:capacity}).

Capacity helps when it changes admission. In the B200 experiment with
GLM-5.3 at a 512K context length, DOP keeps 16 requests resident rather
than 12 for DP-attention.
 The scheduler must weigh
extra capacity against its recurring activation exchange.

\subsection{Switch primitive costs}
\label{sec:switch-microbench}

\begin{figure}[t]
\centering
\includegraphics[width=\linewidth]{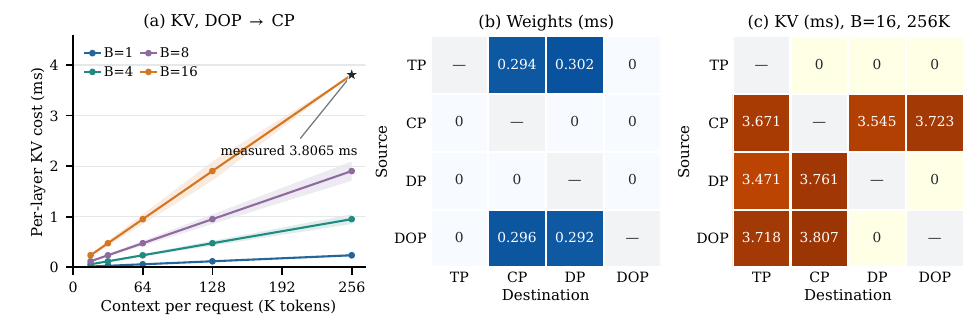}
\caption{Per-layer network cost of the switch primitives on B200.
(a) KV cost of DOP$\to$CP versus context, scaled linearly in $Bs$ from
 the measured point (star); bands show approximately $\pm10\%$ observed
 variation across 20 runs per case.
(b) Weight cost of the twelve switches; rows are sources, columns
 destinations, and 0 marks no network transfer.
(c) KV cost at $B=16$ and 256K. DP denotes DP-attention.}
\label{fig:hotswitch-preparation}
\label{fig:hotswitch-operators}
\end{figure}

Every switch runs one discard, all-gather, or all-to-all per state and layer
(\autoref{fig:splash}, right), so we time these primitives one layer at a
time on B200 with inference paused. A discard has zero network cost;
\autoref{fig:hotswitch-preparation} reports the two collectives.

\paragraph{Weights cost a fixed amount.}
The four weight all-gathers, from TP or DOP into DP-attention or
$\mathrm{CP}$, each move $(1-1/T)$ of one layer's projections
per rank and take 0.292--0.302\,ms per layer
(\autoref{fig:hotswitch-preparation}b). Their payload is independent of
batch size and retained context.

\paragraph{KV cost scales with context.}
The seven KV all-gathers and all-to-alls take 3.471--3.807\,ms per layer at
$B=16$ and 256K tokens per request (\autoref{fig:hotswitch-preparation}c).
Scaling linearly in $Bs$, 
the measured 3.8065\,ms of DOP$\to\mathrm{CP}$ gives an estimated 0.238\,ms at 16K for the same batch
(\autoref{fig:hotswitch-preparation}a).

\subsection{Live switching overhead}
\label{sec:live-overhead}

We compare blocking, \textsc{Splash}, and a target-ready reference on B200
(\autoref{tab:hotswitch-live}). Blocking prepares all missing state before
destination execution; \textsc{Splash} overlaps preparation with computation;
target-ready starts with resident destination state. 
We report the measured
times $T$ and overheads $\Delta T$ over target-ready;
\autoref{app:hotswitch-protocol} gives the timing protocol.

\begin{table}[t]
\caption{All twelve directed switches on B200 ($B=16$, 256K context
tokens per request). Times are milliseconds: 
$T$ is the measured execution-time p50;
$\Delta T$ is p50(p95) overhead over target-ready (Ready). Block. is blocking;
DP is DP-attention.}
\label{tab:hotswitch-live}
\centering\fontsize{7.2}{9}\selectfont
\setlength{\tabcolsep}{1.1pt}
\begin{adjustbox}{max width=\linewidth}
\begin{tabular}{@{}lrrrrr@{\hspace{6pt}}lrrrrr@{}}
\toprule
 & \multicolumn{3}{c}{$T$, p50} & \multicolumn{2}{c}{$\Delta T$, p50(p95)}
 & & \multicolumn{3}{c}{$T$, p50} & \multicolumn{2}{c}{$\Delta T$, p50(p95)} \\
\cmidrule(lr){2-4}\cmidrule(lr){5-6}\cmidrule(lr){8-10}\cmidrule(lr){11-12}
Direction & Ready & Block. & \textsc{Splash} & Block. & \textsc{Splash}
 & Direction & Ready & Block. & \textsc{Splash} & Block. & \textsc{Splash} \\
\midrule
$\mathrm{TP}\!\to\!\mathrm{CP}$ & 2239.33 & 2299.43 & 2240.16 & 60.10(113.03) & 0.83(1.43) & $\mathrm{DP}\!\to\!\mathrm{TP}$ & 5404.63 & 6009.10 & 5412.74 & 604.47(1030.55) & 8.10(13.50) \\
$\mathrm{TP}\!\to\!\mathrm{DP}$ & 2452.25 & 2510.73 & 2453.10 & 58.48(108.61) & 0.85(1.47) & $\mathrm{DP}\!\to\!\mathrm{CP}$ & 2244.30 & 2855.93 & 2252.44 & 611.63(1035.20) & 8.14(14.13) \\
$\mathrm{TP}\!\to\!\mathrm{DOP}$ & 2011.17 & 2011.20 & 2011.19 & 0.03(0.07) & 0.02(0.07) & $\mathrm{DP}\!\to\!\mathrm{DOP}$ & 1964.62 & 1964.64 & 1964.64 & 0.02(0.10) & 0.02(0.08) \\
$\mathrm{CP}\!\to\!\mathrm{TP}$ & 5526.08 & 6132.62 & 5533.86 & 606.53(1045.79) & 7.78(13.67) & $\mathrm{DOP}\!\to\!\mathrm{TP}$ & 5432.56 & 6053.79 & 5442.33 & 621.23(1042.87) & 9.77(13.97) \\
$\mathrm{CP}\!\to\!\mathrm{DP}$ & 2461.12 & 3096.17 & 2469.01 & 635.05(1030.55) & 7.89(13.29) & $\mathrm{DOP}\!\to\!\mathrm{CP}$ & 2346.75 & 3014.06 & 2358.50 & 667.31(1298.42) & 11.76(16.79) \\
$\mathrm{CP}\!\to\!\mathrm{DOP}$ & 2034.72 & 2664.64 & 2042.40 & 629.92(1031.19) & 7.69(13.53) & $\mathrm{DOP}\!\to\!\mathrm{DP}$ & 2450.66 & 2511.54 & 2451.43 & 60.89(107.41) & 0.78(1.39) \\
\bottomrule
\end{tabular}
\end{adjustbox}
\end{table}

\paragraph{Exposed overhead.}
Across all twelve directions, \textsc{Splash} adds 0.02--11.76\,ms at p50
and 0.07--16.79\,ms at p95. Its median overhead stays below 0.51\% of the
corresponding target-ready median, whose latency ranges from 1.96 to
5.53\,s. 
That latency is one prefill step of the destination layout, ordered as in
\autoref{fig:regime-crossover}a at 256K; under chunked prefill a 256K prompt
spans 16--128 such steps, so against time to first token the overhead is
smaller still. For the ten directions requiring network transfers,
\textsc{Splash} reduces median switching overhead by 98.24--98.78\%
relative to blocking. Each overhead is close to one layer's share of the
blocking cost: CP$\to$TP adds 7.78\,ms against 606.53\,ms for all 78 layers,
so the transfers of the other layers hide behind computation.

\paragraph{The missing state determines the remaining cost.}
Weight-only switches add 0.78--0.85\,ms at the median, while KV-only
switches add 7.69--9.77\,ms. DOP$\to\mathrm{CP}$ transfers
both weights and KV and has the largest overhead: 11.76\,ms at p50 and
16.79\,ms at p95, compared with 667.31 and 1,298.42\,ms for blocking.
TP$\to$DOP and DP-attention$\to$DOP 
reuse resident state and add 0.02\,ms at the median, and 0.03 and 0.02\,ms
when blocking, respectively. 
The three switches into TP are measured for their overhead; \textsc{Splash}
rejects any switch that would overflow the destination's memory, as a switch
into TP would at this size (\autoref{eq:feasible}), and appendix
\autoref{tab:capacity} lists the KV capacity each layout supports. These costs support
a small, direction-dependent switching margin for this workload.

\section{Related Work}
\label{sec:related-work}

\paragraph{Live parallelism reconfiguration.}
LoongServe changes sequence parallelism for long-context serving
\citep{loongserve}. Flying Serving switches DP and TP with reusable weight
views and a KV adaptor, and Shift Parallelism alternates TP and sequence
parallelism over compatible KV layouts \citep{flyingserving,shiftpar}.
Moebius reconfigures TP/EP for MoE serving, and ReMP and PipeLive reconfigure
model and pipeline parallelism at runtime \citep{moebius,remp,pipelive}.
Their switch cost depends on which state source and target share, reaching a few hundred milliseconds for Moebius and seconds for ReMP. \textsc{Splash} makes this dependence explicit for attention inside a fixed worker group: it separates weight and KV placement, adds DOP, 
and prices the missing state of each switch as discards, all-gathers, and all-to-alls.

\paragraph{Parallel attention and serving schedules.}
Megatron-LM established tensor-parallel projections \citep{megatronlm2019,megatronlm}; Ring
Attention, Ulysses, and inference CP distribute within-context work
\citep{ringattention,ulysses,cp-million}, and SGLang provides DP-attention
and several CP storage paths \citep{sglang,sglangv04,sglang-cp-roadmap,zai-layersplit}.
\citet{palm-inference} shard multiquery attention over the batch on TPUs to
avoid replicating its KV head; DOP removes the projection
replicas that DP-attention keeps for absorbed MLA.
Orca and Sarathi-Serve schedule iterations and prefill/decode batches
\citep{orca,sarathi}; \textsc{Splash} chooses the attention layout beneath such
schedules.

\paragraph{Cache representation and disaggregation.}
MQA, GQA, and MLA shrink stored state \citep{mqa,gqa,deepseekv2,deepseekv3};
PagedAttention, quantization, and eviction manage what remains
\citep{pagedattention,kivi,h2o}, and FlashAttention and FlashInfer speed up
the kernels \citep{flashattention,flashinfer}. These change \textsc{Splash}'s profiles,
not its placement decision; sparse and hybrid attention add indexer or
recurrent state, which the handoff moves with each layer's KV \citep{nsa,deepseekv32,kimilinear}.
DistServe, Splitwise, and Mooncake separate prefill from decode and organize
serving around KV transport \citep{distserve,splitwise,mooncake}; 
\textsc{Splash} gives each phase its own layout, and its switches share bandwidth with prefill-to-decode KV transfer.

\section{Discussion and Conclusion}
\label{sec:conclusion}

Attention layout is a choice that can change within a request's lifetime.
Viewing a layout as two ownership decisions, where the projections live and
who owns each request's history, reduces every one of the twelve switches to
discards, all-gathers, and all-to-alls, and exposes DOP, which stores each
projection and each history once. The evaluation confirms both consequences.
On B200 GPUs serving GLM-5.3, each of the four layouts leads a regime of
context length and concurrency, and following the best one improves
end-to-end throughput by 1.3--1.73$\times$ over fixed-layout deployments; DOP
leads at 256K--512K and at the largest batches, by 3.4--11.0\% over the
strongest alternative, and holds 27.3\% more KV than DP-attention. The same
regimes appear on H200 and DCU. Moving between regimes is cheap: one layer's
communication takes 0.29--0.30\,ms for weights and 3.47--3.81\,ms for KV at
$B=16$ and 256K, and because later layers hide behind computation, a complete
switch adds 0.02--11.76\,ms end to end, 98.24--98.78\% less than a blocking
switch wherever state crosses the network. Workloads whose batches narrow from
many short requests to a few long ones, such as RL rollout, can therefore let
the layout follow the batch, and choosing batch size and admission jointly
with the layout is the natural next step. \autoref{app:limitations} details
the remaining costs.

\subsubsection*{AI Use Statement}
LLM-based tools assisted with coding, experiments and writing. The authors reviewed
all AI-assisted work and take full responsibility for the content of this paper.

\bibliography{references}
\bibliographystyle{preprint}

\appendix

\section{Weight Transfer Accounting}
\label{app:weights}

Let $S_{A,k}$ be one complete set of layer-$k$ attention projections whose
placement changes, in the transfer format. Count each independent parameter
once, including required quantization scales. A resident derived kernel copy
does not require a second network transfer if it can be reconstructed locally;
its materialization time and memory still count. Aggregate traffic is
$V_{\mathrm{sum},k}^W=\sum_r V_{r,k}^W$, counting each receive once rather
than adding sends and receives or multiplying by fabric hops.

TP and DOP use the same projection shards; DP-attention and CP
replicate those projections. Thus only TP or DOP switching into
DP-attention or CP receives $(1-1/T)S_{A,k}$ weight bytes per rank.
The other directions reuse resident weights.

For standard MLA with low-rank queries, the changed matrices are Q-B, KV-B,
and O. Q-A, KV-A, rotary-key rows, and normalization parameters already
replicated in the implementation are excluded. For hidden width $d$,
$H$ heads, query rank $r_q$, latent width $d_c$, and per-head non-rotary,
rotary, and value widths $d_n,d_r,d_v$,
\begin{equation}
 S_{A,k}=
 H\big[b_Qr_q(d_n+d_r)+b_{KV}d_c(d_n+d_v)+b_Od\,d_v\big]
 +S_{\mathrm{meta},k}.
 \label{eq:mla-weight-bytes}
\end{equation}
The $b$ terms specify bytes per parameter in each transfer format.
$S_{\mathrm{meta},k}$ accounts for scales and other required metadata.

DeepSeek-V3 uses
$(d,H,r_q,d_c,d_n,d_r,d_v)=(7168,128,1536,512,128,64,128)$
\citep{deepseekv3}. Q-B, KV-B, and O therefore occupy 72, 32, and
224\,MiB in BF16, totaling $328$\,MiB per layer. At $T=8$, a
sharded-to-replicated step receives $287$\,MiB per rank, or
$2{,}296$\,MiB across the group. Uniform FP8 halves the parameter payload
to $143.5$\,MiB per rank before scales. 
The same count gives 145.73\,MiB per rank per layer for GLM-5.3 in FP8 at
$T=8$ (\autoref{sec:prepare}). These are analytical sizes, not measured
migration times, and measured memory differs from them because the runtime
also allocates beyond the counted projection bytes: on B200, DOP's per-GPU
KV pool exceeds DP-attention's by 12.69\,GiB (\autoref{tab:capacity}), the
measured value \autoref{sec:dop} reports.

\section{Per-Step Cost Model}
\label{app:inference-model}
\label{sec:inference-bottlenecks}

To isolate the placement tradeoff, consider dense MLA inference on the same
$T$ devices, with $B$ equal-length requests and $q$ query tokens processed
per request in an iteration. Let $s$ be the live context length including
the current chunk, with $1\leq q\leq s$, and $N=Bq$ the current query rows.
In this comparison, TP and DOP shard the
large attention projections, whereas DP-attention and CP replicate them.
TP replicates the shared latent KV; the other three layouts store each KV
token once across the group. Request ownership and context partitioning
are different ways to realize that unreplicated storage.
\autoref{tab:rank-loads} lists the resulting busiest-rank loads.

\begin{table}[htbp]
\caption{Busiest-rank load for $B$ equal-length requests on $T$ ranks,
with count-balanced request ownership and ideal token partitioning for CP.
KV stored and KV read count histories of $ks$ bytes, and
attention work counts one request's attention.
\autoref{eq:inference-bottleneck-cost} converts these loads into time. The owner
entries equal $B/T$ whenever $T$ divides $B$, matching
\autoref{eq:inference-layout-memory}.}
\label{tab:rank-loads}
\centering\small
\begin{tabular}{lcccc}
\toprule
Layout & Weights & KV stored & KV read & Attention work \\
\midrule
TP & $W_A/T$ & $B$ & $B$ & $B/T$ \\
CP & $W_A$ & $B/T$ & $B/T$ & $B/T$ \\
DP-attention & $W_A$ & $\lceil B/T\rceil$ & $\lceil B/T\rceil$ & $\lceil B/T\rceil$ \\
DOP & $W_A/T$ & $\lceil B/T\rceil$ & $\lceil B/T\rceil$ & $\lceil B/T\rceil$ \\
\bottomrule
\end{tabular}
\end{table}

The relevant bottleneck need not stay fixed as $s$ grows. For layout
$\ell$, a stage-wise cost model is
\begin{equation}
\begin{aligned}
 t^{\mathrm{proj}}_\ell(q)
 &=\max\left\{\frac{F^{\mathrm{proj}}_\ell(q)}
                        {\Phi^{\mathrm{proj}}_\ell},
                 \frac{W_\ell}{\beta^W_\ell}\right\},\\
 t^{\mathrm{att}}_\ell(s,q)
 &=\max\left\{\frac{F^{\mathrm{att}}_\ell(s,q)}
                        {\Phi^{\mathrm{att}}_\ell(s,q)},
                 \frac{K^{\mathrm{read}}_\ell(s,q)}
                      {\beta^{\mathrm{KV}}_\ell(s,q)}\right\},\\
 t_\ell(s,q)
 &=t_0+t^{\mathrm{proj}}_\ell(q)+t^{\mathrm{att}}_\ell(s,q)
   +\delta_\ell
   +\left[R_\ell(s,q)-\eta_\ell t^{\mathrm{att}}_\ell(s,q)\right]_+ .
\end{aligned}
\label{eq:inference-bottleneck-cost}
\end{equation}
Here $F$ denotes per-rank arithmetic work, $W_\ell$ resident projection
weights, $K^{\mathrm{read}}_\ell$ KV bytes read, and $\Phi$ and $\beta$
effective compute and memory rates. The communication bandwidth cost
$R_\ell$ depends on the communication scheme and query schedule. It can
overlap only with the explicitly budgeted fraction
$\eta_\ell\in[0,1]$ of attention work; collective latency $\delta_\ell$
remains exposed. Such overlap requires chunked or cross-request
pipelining and sufficient workspace. Projection and attention costs are
added because they are dependent stages, rather than treating all work
as one perfectly overlapping resource maximum.

At fixed $q$, dense attention work and the minimum history-read volume
both grow approximately linearly with $s$. Longer histories can therefore
shift the dominant cost away from weight reads or fixed communication
toward attention computation or KV bandwidth. Length alone does not force
a change between the latter two limits: their asymptotic ratio is constant
when effective rates are fixed. Length-dependent utilization, tiling, and
communication overlap can nevertheless change realized costs and produce
crossovers. DP-attention and DOP have the same owner-local core work under
equal load balance; their projection and redistribution costs differ. CP
exposes a different work decomposition and additional cooperation. No
pairwise crossing is guaranteed by a layout name alone.

\subsection{Theoretical Scenarios for DCU and H200}
\label{app:theory-hardware}

\begin{figure}[htbp]
\centering
\includegraphics[width=\linewidth]{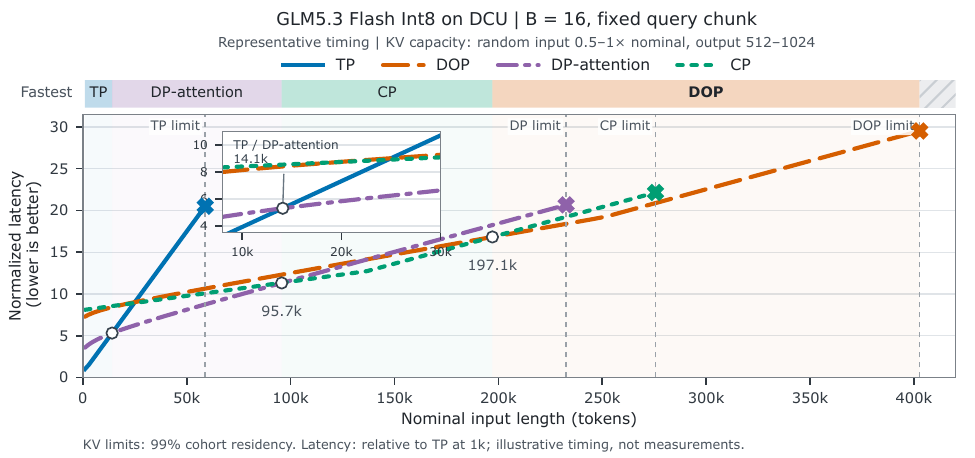}
\caption{Conditional latency model for GLM-5.3-Flash INT8 on BW1000 DCUs
($B=16$, fixed query chunk), normalized to TP at 1k. Open circles mark
changes in the fastest feasible layout; the inset enlarges the first.
Crosses mark KV capacity limits, and hatched regions indicate where the KV cache exceeds the available capacity. \autoref{app:inference-model}
gives the timing and capacity assumptions.}
\label{fig:theory-dcu-reference}
\end{figure}

\begin{figure}[htbp]
\centering
\includegraphics[width=\linewidth]{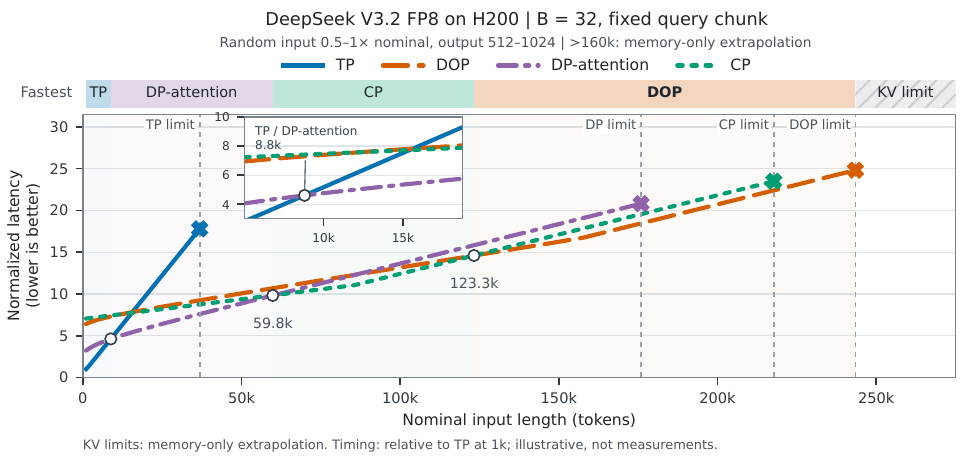}
\caption{Conditional latency model for DeepSeek-V3.2 FP8 on H200 GPUs
($B=32$, fixed query chunk), normalized to TP at 1k. Open circles mark
changes in the fastest feasible layout; the inset enlarges the first.
Crosses mark KV capacity limits, and hatched regions indicate where the KV cache exceeds the available capacity. \autoref{app:inference-model}
gives the timing and capacity assumptions.}
\label{fig:theory-deepseek-h200}
\end{figure}

\autoref{fig:theory-dcu-reference} and \autoref{fig:theory-deepseek-h200}
provide the DCU reference and H200 counterpart to the B200 illustration in
\autoref{fig:inference-bottleneck-latency}. They use a fixed query chunk and
an illustrative timing law, separate from the scheduler's deployment-specific
calibration. The DCU analysis considers GLM-5.3-Flash on BW1000 accelerators.
The H200 scenario uses DeepSeek-V3.2 FP8 on H200 GPUs with $B=32$.

\paragraph{Shared timing assumptions.}
Let $x=L/1024$ for nominal input length $L$, and
$a(x)=0.05x+1.2x/(x+8)$. The dimensionless reference costs are
\begin{equation}
\begin{aligned}
 \tau_{\mathrm{TP}}(x)&=0.40+0.25x+[0.15-0.0625x]_+,\\
 \tau_{\mathrm{DP}}(x)&=2.45+a(x),\\
 \tau_{\mathrm{CP}}(x)&=2.55+0.05x+[3.40-0.025x]_+,\\
 \tau_{\mathrm{DOP}}(x)&=0.45+a(x)+[4.80-0.35a(x)]_+.
\end{aligned}
\label{eq:theory-reference-costs}
\end{equation}
These are one parameterization of \autoref{eq:inference-bottleneck-cost};
both owner layouts use the same core time $a(x)$. The coefficients specify
assumed utilization, communication, and overlap, rather than measured rates.
The plotted latency in scenario $h$ is
\begin{equation}
 \widetilde t_{h,\ell}(x)=
 \frac{\tau_\ell(x/\alpha_h)}{\tau_{\mathrm{TP}}(1/\alpha_h)},
 \qquad
 \alpha_{\mathrm{DCU}}=1,\quad
 \alpha_{\mathrm{H200}}=0.6252985669,\quad
 \alpha_{\mathrm{B200}}=1.2853801752.
\label{eq:theory-capacity-scaling}
\end{equation}
The H200 and B200 factors are their sampled TP input limits divided by
the reference limit of 58.875k. This capacity-based scaling changes all four
curves together; it does not calibrate hardware speed, sparse attention,
indexer execution, or batch-dependent kernel efficiency. Each figure is
normalized independently to TP at 1k, so its vertical values do not compare
absolute hardware speeds or full-request latency.

\paragraph{Residency assumptions.}
Inputs are sampled independently and uniformly from the integers in
$[\lfloor L/2\rfloor,L]$, with independent output lengths in $[512,1024]$.
The entire input and requested output KV are reserved. DP-attention and DOP
use count-balanced, length-blind round-robin owners; CP ideally balances
stored tokens. The limits use 200,000 cohorts with PCG64 seed 20260922
and a 99\% all-rank residency target. The timing law is representative at
nominal $L$, not a mean or percentile over these random cohorts.

The DCU reference uses $B=16$, $T=8$, 52.8\,GiB per rank for attention
weights and latent KV, a separate 1\,GiB transient reserve, 24\,GiB of full
attention weights, and 64\,KiB of KV per token across layers. Its sampled
nominal-input limits are 58.9k, 232.6k, 275.6k, and 402.8k for TP,
DP-attention, CP, and DOP, respectively ($1\mathrm{k}=1024$ tokens).

For DeepSeek-V3.2 FP8 on H200, the memory fraction is 0.90 and the
61-layer KV payload is 48,068 bytes/token, including FP8 latent and
indexer state, their scales, and BF16 RoPE. The per-GPU token pools in
TP, DP-attention, CP, DOP order are 1,000,832, 686,144, 724,736, and
948,352. They are analytical projections from GLM-5.3 H200 allocations,
replacing parameter and KV payloads while keeping the non-parameter
reserve fixed, not measurements of DeepSeek initialization. At $B=32$,
the corresponding sampled nominal-input limits are 36.81k, 175.93k,
217.73k, and 243.48k.

The B200 illustration in \autoref{fig:inference-bottleneck-latency} uses
GLM-5.3 FP8, $B=16$, and memory fraction 0.85. Its displayed TP limit is
the sampled 75.68k boundary; the DP-attention, CP, and DOP limits are the
prescribed 400k, 450k, and 512k scenario boundaries, respectively, rather
than the same 99\% target. Across these figures, changes in model, batch,
and memory allocation accompany the hardware change. The resulting
crossovers are conditional illustrations, not measured switching thresholds.

\section{Experimental Configuration}
\label{app:repro}

\paragraph{Code availability.}
Our implementation is open source at\\
\url{https://github.com/ict-agent/SPLASH-sglang}.

We ran the B200 and H200 experiments on GPU servers rented from Vast.ai
(\url{https://vast.ai}); \autoref{app:b200-config} and
\autoref{app:h200-config} give their configurations.
All DCU experiments use BW1000 accelerators.

\subsection{B200 fixed-layout sweeps}
\label{app:b200-config}

The experiments use GLM-5.3 with 78 layers, FP8 weights, BF16 activations,
and FP8 E4M3 KV on \textsc{Splash}, built atop SGLang 0.5.10 with the sparse
FlashMLA backend. The model's nominal context limit is 1,048,576 tokens, and the
runtime limit is 525,376 tokens.

Prefill runs on B200 GPUs on one node with TP8/EP8 and memory fraction
0.85.
DP-attention and DOP use attention-DP8. Decode runs on sixteen
GPUs across two nodes with TP16/DP16/EP16.

Four sweeps compare the prefill layouts at eight input lengths from 1K
to 512K with $B=16,32$, and eight batch sizes from 1 to 256 at 64K and
128K input. Each request generates 1,024 tokens, and K denotes 1,024 tokens.
Each point averages the batch end-to-end time $T_{\mathrm{batch}}$ over 10
runs; $T_{\mathrm{batch}}$ includes queueing, prefill, KV transfer, and decode. Input construction,
cache flushing, and metric collection are outside that interval.
Total throughput is $B(L_{\mathrm{in}}+1{,}024)/T_{\mathrm{batch}}$;
\autoref{tab:full-sweep} lists all 128 measurements.

\subsection{DCU fixed-layout sweeps}
\label{app:dcu-config}

The experiments use GLM-5.3-Flash 300B with INT8 weights, FP8 KV,
and 45 layers (11 MLA and 34 KDA) in SGLang's language-only path.
Prefill runs on DCUs on one node; decode runs on sixteen DCUs
across two nodes with TP16/DP16/EP16.

Prefill and decode use memory fraction 0.85.
\autoref{tab:capacity} reports the KV pools and effective capacity.

Four sweeps compare the prefill layouts at eight input lengths from 1K
to 512K with $B=16,32$, and nine batch sizes from 1 to 512 at 32K and
64K input. Each request generates 1,024 tokens. 
Each point averages the batch end-to-end time over 10 runs; that time
includes queueing, prefill, KV transfer, and decode, while input construction,
cache flushing, and metric collection are outside it.
Total throughput uses the same definition as the B200 sweeps;
\autoref{tab:dcu-full-sweep} lists all 136 measurements.

\subsection{H200 fixed-layout sweeps}
\label{app:h200-config}

The experiments use DeepSeek-V3.2 on H200 with one prefill instance
and one decode instance (1P1D). We compare TP, CP, DP-attention,
and DOP as prefill layouts.

Four sweeps cover seven input lengths from 1K to 256K at $B=16,32$,
eight batch sizes from 1 to 256 at 32K input, and seven batch sizes
from 1 to 128 at 64K input. Each request generates 1,024 tokens.

\begin{figure}[htbp]
\centering
\includegraphics[width=\linewidth]{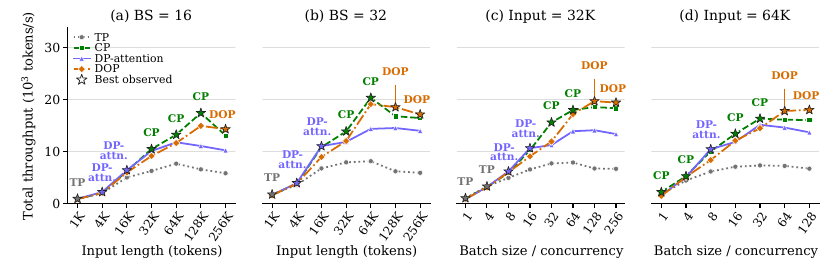}
\caption{Fixed-layout DeepSeek-V3.2 performance on H200 with 1P1D
and 1K output tokens per request.
(a, b) Input-length sweeps at $B=16,32$; (c, d) batch-size sweeps at
32K and 64K input. Sampled positions are equally spaced.
Stars and labels mark the best recorded layout at each position.}
\label{fig:h200-appendix-regime-crossover}
\label{fig:h200-appendix-regime-batch}
\end{figure}

Each point averages the batch end-to-end time over 10 runs. Total throughput uses
the same definition as the B200 and DCU sweeps.
\autoref{fig:h200-appendix-regime-crossover} plots the results, and
\autoref{tab:h200-full-sweep} lists all 116 measurements.

\subsection{KV capacity and DCU admission}

The long-context mixed-serving experiment uses GLM-5.3-Flash with
INT8 weights and FP8 KV on DCUs at memory fraction 0.85.
The 240,000/1,024 comparison at 32 requests has three confirmation runs
after warmup. These co-located serving experiments complement the
prefill/decode-disaggregated sweeps.

\begin{table}[t]
\caption{KV pools and effective capacity for accelerator groups.
The platforms use different models: GLM-5.3 on B200, DeepSeek-V3.2 on H200,
and GLM-5.3-Flash on BW1000 DCUs. Memory fraction is the configured
allocation fraction. Pool sizes are in GiB; effective capacity counts
distinct tokens across the group, excluding replicated copies.
Displayed GiB values are rounded.}
\label{tab:capacity}
\centering\small
\setlength{\tabcolsep}{4pt}
\begin{tabular}{llrrr}
\toprule
Device / fraction & Layout & GiB/card & GiB/group & Distinct tokens \\
\midrule
B200 / 0.85 & TP & 59.84 & 478.72 & 1,045,312 \\
 & DOP & 59.19 & 473.51 & 8,271,360 \\
 & CP & 48.22 & 385.79 & 6,738,944 \\
 & DP-attention & 46.50 & 371.98 & 6,497,792 \\
\midrule
H200 / 0.90 & TP & 36.21 & 289.68 & 632,512 \\
 & DOP & 33.86 & 270.90 & 4,731,904 \\
 & CP & 22.93 & 183.47 & 3,204,608 \\
 & DP-attention & 21.21 & 169.67 & 2,963,456 \\
\midrule
DCU / 0.85 & TP & 10.27 & 82.16 & 1,786,816 \\
 & DOP & 6.43 & 51.44 & 8,949,248 \\
 & CP & 5.70 & 45.60 & 7,933,236 \\
 & DP-attention & 5.40 & 43.23 & 7,521,792 \\
\bottomrule
\end{tabular}
\end{table}

\paragraph{Admission confirmation.}
A three-run confirmation of the admission effect in \autoref{app:dcu-capacity},
at 240K input and 1K output tokens with 32 requests, gives DOP a
$1.322\times$ burst-throughput gain over DP-attention (14,846.29 versus
11,228.56 input tokens/s) at a higher mean completion latency (516.84 versus
477.09\,s).

\section{Complete Fixed-Layout Results}
\label{app:full-sweep}

\subsection{B200 results}
\autoref{tab:full-sweep} reports all 128 measurements used by
\autoref{fig:regime-crossover}. Values are total throughput in
$10^3$ tokens/s; bold identifies the largest recorded value at each
position. Overlapping shapes retain independently recorded measurements,
and each configuration averages 10 runs.

\begin{table}[htbp]
\caption{B200 fixed-layout sweeps, in total throughput ($10^3$ tokens/s).
Each row has $B$ requests at client concurrency $B$ and 1,024 output
tokens per request. Input lengths are exact token counts.
DP-attn.\ abbreviates DP-attention.}
\label{tab:full-sweep}
\centering\small
\begin{tabular}{lrrrrr}
\toprule
Input & $B$ & TP & CP & DP-attn. & DOP \\
\midrule
\multicolumn{6}{l}{Input-length sweep, $B=16$} \\
1,024 & 16 & \textbf{1.529} & 1.176 & 1.000 & 1.053 \\
4,096 & 16 & \textbf{3.285} & 3.188 & 3.110 & 3.111 \\
16,384 & 16 & 7.362 & 8.539 & \textbf{9.380} & 8.738 \\
32,768 & 16 & 9.279 & 13.723 & \textbf{13.910} & 12.594 \\
65,536 & 16 & 11.140 & 17.806 & \textbf{18.424} & 16.287 \\
131,072 & 16 & 11.647 & \textbf{22.757} & 21.739 & 20.650 \\
262,144 & 16 & 10.955 & 22.149 & 19.096 & \textbf{22.903} \\
524,288 & 16 & 8.913 & 20.091 & 17.857 & \textbf{22.104} \\
\midrule
\multicolumn{6}{l}{Input-length sweep, $B=32$} \\
1,024 & 32 & \textbf{2.835} & 2.675 & 2.428 & 2.580 \\
4,096 & 32 & 5.241 & 5.855 & \textbf{6.234} & 5.752 \\
16,384 & 32 & 9.861 & 13.804 & \textbf{14.111} & 12.762 \\
32,768 & 32 & 11.595 & 18.409 & \textbf{18.983} & 16.504 \\
65,536 & 32 & 12.480 & \textbf{25.046} & 22.636 & 24.382 \\
131,072 & 32 & 12.398 & \textbf{25.786} & 21.105 & 23.741 \\
262,144 & 32 & 11.285 & 21.221 & 19.819 & \textbf{23.470} \\
524,288 & 32 & 9.039 & 20.173 & 19.128 & \textbf{21.059} \\
\midrule
\multicolumn{6}{l}{Batch-size sweep, 64K input} \\
65,536 & 1 & \textbf{2.716} & 2.601 & 2.387 & 2.120 \\
65,536 & 4 & 6.775 & 7.307 & \textbf{7.913} & 7.041 \\
65,536 & 8 & 9.465 & 13.411 & \textbf{13.451} & 12.240 \\
65,536 & 16 & 11.347 & 18.291 & \textbf{18.438} & 16.236 \\
65,536 & 32 & 12.592 & \textbf{25.783} & 20.857 & 23.634 \\
65,536 & 64 & 13.327 & \textbf{24.928} & 19.642 & 23.681 \\
65,536 & 128 & 13.727 & 24.767 & 19.332 & \textbf{26.812} \\
65,536 & 256 & 13.920 & 24.669 & 18.960 & \textbf{26.822} \\
\midrule
\multicolumn{6}{l}{Batch-size sweep, 128K input} \\
131,072 & 1 & 4.330 & \textbf{5.514} & 2.913 & 3.189 \\
131,072 & 4 & 9.011 & \textbf{14.061} & 8.771 & 9.123 \\
131,072 & 8 & 10.603 & 13.644 & \textbf{14.613} & 14.196 \\
131,072 & 16 & 11.873 & \textbf{23.128} & 22.534 & 20.571 \\
131,072 & 32 & 12.633 & \textbf{25.600} & 23.010 & 23.250 \\
131,072 & 64 & 13.053 & \textbf{27.376} & 23.105 & 24.838 \\
131,072 & 128 & 13.263 & 27.709 & 24.622 & \textbf{30.281} \\
131,072 & 256 & 13.526 & 27.677 & 25.044 & \textbf{30.724} \\
\bottomrule
\end{tabular}
\end{table}

\subsection{DCU results}

\autoref{tab:dcu-full-sweep} contains the 136 layout measurements used in
\autoref{fig:dcu-appendix-regime-crossover}. Bold identifies
the largest observed total throughput at each sampled position. Values are
in $10^3$ tokens/s and rounded to three decimals. 
Each measurement averages 10 runs. The repeated
32K and 64K shapes at $B=16,32$ retain their respective sweep measurements.

\begin{table}[htbp]
\caption{Updated DCU fixed-layout sweeps, in total throughput
($10^3$ tokens/s). Each row has $B$ requests at client concurrency $B$ and
1,024 output tokens per request. Input lengths are exact token counts.
DP-attn.\ abbreviates DP-attention.}
\label{tab:dcu-full-sweep}
\centering\small
\begin{tabular}{lrrrrr}
\toprule
Input & $B$ & TP & CP & DP-attn. & DOP \\
\midrule
\multicolumn{6}{l}{Input-length sweep, $B=16$} \\
1,024 & 16 & \textbf{1.234} & 1.139 & 0.985 & 1.110 \\
4,096 & 16 & \textbf{2.706} & 2.632 & 2.339 & 2.593 \\
16,384 & 16 & 5.301 & 7.005 & \textbf{7.337} & 7.122 \\
32,768 & 16 & 6.999 & 10.559 & \textbf{11.515} & 10.989 \\
65,536 & 16 & 8.366 & 14.283 & \textbf{16.205} & 15.248 \\
131,072 & 16 & 9.122 & \textbf{17.534} & 15.776 & 16.249 \\
262,144 & 16 & 8.421 & 19.558 & 18.690 & \textbf{21.539} \\
524,288 & 16 & 8.405 & 20.138 & 18.790 & \textbf{22.338} \\
\midrule
\multicolumn{6}{l}{Input-length sweep, $B=32$} \\
1,024 & 32 & \textbf{2.051} & 2.023 & 1.985 & 1.941 \\
4,096 & 32 & 3.855 & 4.463 & \textbf{4.560} & 4.468 \\
16,384 & 32 & 7.150 & 10.526 & \textbf{11.369} & 10.979 \\
32,768 & 32 & 8.603 & 14.445 & \textbf{16.353} & 15.385 \\
65,536 & 32 & 9.466 & \textbf{17.833} & 16.101 & 15.255 \\
131,072 & 32 & 9.750 & \textbf{19.990} & 18.369 & 18.763 \\
262,144 & 32 & 9.439 & 21.027 & 20.031 & \textbf{23.436} \\
524,288 & 32 & 8.552 & 20.941 & 20.750 & \textbf{23.324} \\
\midrule
\multicolumn{6}{l}{Batch-size sweep, 32K input} \\
32,768 & 1 & \textbf{1.036} & 0.943 & 0.921 & 0.914 \\
32,768 & 4 & 3.259 & \textbf{3.864} & 3.603 & 3.569 \\
32,768 & 8 & 5.073 & 6.691 & \textbf{6.955} & 6.739 \\
32,768 & 16 & 7.005 & 10.574 & \textbf{11.517} & 10.982 \\
32,768 & 32 & 8.619 & 14.404 & \textbf{16.274} & 15.273 \\
32,768 & 64 & 9.739 & \textbf{18.056} & 16.172 & 17.773 \\
32,768 & 128 & 10.447 & \textbf{20.545} & 19.148 & 20.059 \\
32,768 & 256 & 10.813 & 22.093 & 20.656 & \textbf{24.801} \\
32,768 & 512 & 11.534 & 22.932 & 22.255 & \textbf{25.991} \\
\midrule
\multicolumn{6}{l}{Batch-size sweep, 64K input} \\
65,536 & 1 & 1.870 & \textbf{2.062} & 1.581 & 1.559 \\
65,536 & 4 & 4.933 & \textbf{6.537} & 5.918 & 5.774 \\
65,536 & 8 & 6.794 & 10.263 & \textbf{11.107} & 10.582 \\
65,536 & 16 & 8.371 & 14.318 & \textbf{16.219} & 15.354 \\
65,536 & 32 & 9.468 & 17.705 & \textbf{20.832} & 19.401 \\
65,536 & 64 & 10.120 & \textbf{20.196} & 18.452 & 18.112 \\
65,536 & 128 & 10.498 & 21.802 & 20.290 & \textbf{24.383} \\
65,536 & 256 & 11.360 & 22.623 & 21.902 & \textbf{25.563} \\
65,536 & 512 & 11.360 & 23.036 & 21.474 & \textbf{26.178} \\
\bottomrule
\end{tabular}
\end{table}

\subsection{H200 results}
\label{app:h200-full-results}

\autoref{tab:h200-full-sweep} reports all 116 fixed-layout measurements
for DeepSeek-V3.2 on H200 in a 1P1D deployment. Values are total
throughput in $10^3$ tokens/s, rounded to three decimals; bold identifies
the largest recorded value at each sampled position. 
Each configuration averages 10 runs. Overlapping shapes retain their respective sweep
measurements.

\begin{table}[htbp]
\caption{H200/DeepSeek-V3.2 fixed-layout sweeps, in total throughput
($10^3$ tokens/s). Each row has $B$ requests and 1,024 output tokens
per request. Input lengths are exact token counts.
DP-attn.\ abbreviates DP-attention.}
\label{tab:h200-full-sweep}
\centering\small
\begin{tabular}{lrrrrr}
\toprule
Input & $B$ & TP & CP & DP-attn. & DOP \\
\midrule
\multicolumn{6}{l}{Input-length sweep, $B=16$} \\
1,024 & 16 & \textbf{0.937} & 0.885 & 0.930 & 0.864 \\
4,096 & 16 & 1.994 & 2.080 & \textbf{2.232} & 2.071 \\
16,384 & 16 & 5.037 & 6.267 & \textbf{6.441} & 5.996 \\
32,768 & 16 & 6.300 & \textbf{10.512} & 10.137 & 9.185 \\
65,536 & 16 & 7.700 & \textbf{13.279} & 11.820 & 11.676 \\
131,072 & 16 & 6.596 & \textbf{17.457} & 11.101 & 14.978 \\
262,144 & 16 & 5.847 & 13.119 & 10.265 & \textbf{14.349} \\
\midrule
\multicolumn{6}{l}{Input-length sweep, $B=32$} \\
1,024 & 32 & \textbf{1.769} & 1.680 & 1.667 & 1.598 \\
4,096 & 32 & 3.612 & 3.842 & \textbf{3.964} & 3.867 \\
16,384 & 32 & 6.782 & 10.880 & \textbf{11.069} & 8.985 \\
32,768 & 32 & 7.955 & \textbf{13.913} & 11.900 & 12.064 \\
65,536 & 32 & 8.177 & \textbf{20.386} & 14.402 & 19.170 \\
131,072 & 32 & 6.225 & 16.843 & 14.567 & \textbf{18.584} \\
262,144 & 32 & 5.935 & 16.488 & 14.026 & \textbf{17.186} \\
\midrule
\multicolumn{6}{l}{Batch-size sweep, 32K input} \\
32,768 & 1 & \textbf{1.054} & 0.972 & 0.824 & 0.927 \\
32,768 & 4 & \textbf{3.320} & 3.234 & 3.212 & 3.132 \\
32,768 & 8 & 4.971 & 5.932 & \textbf{6.208} & 6.044 \\
32,768 & 16 & 6.584 & 10.321 & \textbf{10.681} & 9.104 \\
32,768 & 32 & 7.748 & \textbf{15.643} & 11.287 & 11.977 \\
32,768 & 64 & 7.920 & \textbf{18.032} & 13.949 & 17.245 \\
32,768 & 128 & 6.766 & 18.600 & 14.143 & \textbf{19.743} \\
32,768 & 256 & 6.704 & 18.391 & 13.399 & \textbf{19.466} \\
\midrule
\multicolumn{6}{l}{Batch-size sweep, 64K input} \\
65,536 & 1 & 1.784 & \textbf{2.232} & 1.598 & 1.477 \\
65,536 & 4 & 4.450 & \textbf{5.296} & 5.118 & 5.147 \\
65,536 & 8 & 6.190 & 10.050 & \textbf{10.484} & 8.396 \\
65,536 & 16 & 7.109 & \textbf{13.466} & 11.891 & 12.121 \\
65,536 & 32 & 7.365 & \textbf{16.378} & 15.190 & 14.522 \\
65,536 & 64 & 7.261 & 16.169 & 14.651 & \textbf{17.846} \\
65,536 & 128 & 6.734 & 16.103 & 13.736 & \textbf{18.054} \\
\bottomrule
\end{tabular}
\end{table}

\section{DOP Implementation}
\label{app:attention}

CP is a family of execution and storage configurations. Position-sharded
CP partitions one request's token history; layer-split storage partitions
cache-bearing layers; a replicated-cache CP configuration can distribute
prefill work without reducing persistent KV copies. A layer-local token
pool does not represent that many complete model histories independently
on every rank. \textsc{Splash}'s manifest tracks the actual head, position, and
layer blocks so that both residency and migration respect these differences.

The DOP transfer implementation converts a tensor in global-row,
local-head order into owner-row, all-head order and back. It synchronizes
per-owner row counts once for a forward batch and uses variable all-to-all
split sizes for uneven owners. Ranks that own no rows in a batch still enter both
collectives, so mixed and idle batches cannot deadlock. Packing preserves row identity, head order,
and positions. Fused no-position/rotary query packing reduces collective
count. Aligned data can use a 16-byte vectorized path; unaligned data uses
a scalar fallback. The current transfer path requires matching TP and
attention-DP sizes and rank order.

The reverse exchange must preserve all contributions to the sharded output
projection. An all-reduce or equivalent complete reduce-scatter precedes
owner-row selection. Alternatively, an owner-local output GEMM requires
the complete compatible O weights and activations, which is a DOP-P
placement. Tensor-shape checks alone cannot detect an omitted reduction.

For sparse attention, indexer cache, selected-block metadata, and positions
must remain consistent with KV. For recurrent layers, the mutable state
must be handed off after the last source update. Prefix sharing requires
reference-count and copy-on-write handling. CUDA graphs require stable
destination addresses or recapture. These states are part of the handoff contract.

\section{Hot-Switch Measurement Protocol}
\label{app:hotswitch-protocol}

The B200 results in \autoref{sec:switch-microbench} and
\autoref{sec:live-overhead} have separate timing scopes:
single-layer network communication and reported switching overhead.
The following accounting defines the reported quantities and estimates.

\subsection{Communication and workload scaling}
\label{app:hotswitch-communication}

The weight payload is set by projection placement; the logical
latent-KV buffer has size
\begin{equation}
 V_{\mathrm{latent}}=B\,s\,r_{\mathrm{KV}}\,b_{\mathrm{KV}},
 \label{eq:microbench-kv-buffer}
\end{equation}
where $r_{\mathrm{KV}}$ is the latent width and $b_{\mathrm{KV}}$ is
the element size in bytes. Rotary keys, quantization scales, page
metadata, and any sparse or recurrent state need separate inventories.
Logical buffer volume is not the same as per-rank received bytes:
the latter follows the actual state differences in \autoref{eq:kv-missing}.

The B200 KV reference is $B_{\mathrm{ref}}=16$ and
$s_{\mathrm{ref}}=262{,}144$. Holding the representation and transfer
implementation fixed, Figure~\ref{fig:hotswitch-preparation}a uses
\begin{equation}
 \widehat t_K(a\to b;B,s)
 =t_K^{\mathrm{ref}}(a\to b)
   \frac{B}{B_{\mathrm{ref}}}\frac{s}{s_{\mathrm{ref}}},
 \label{eq:microbench-kv-time}
\end{equation}
with the measured DOP$\to$CP reference of 3.8065\,ms.
Each operator case is measured over 20 runs. Shaded bands show the
observed run-to-run variation of approximately $\pm10\%$.

\subsection{Switching overhead}

The target-ready reference begins with the same destination state
already resident. For a matched timed execution interval, define
\begin{equation}
\begin{aligned}
 \Delta T_{\mathrm{blocking}}&=T_{\mathrm{blocking}}-T_{\mathrm{ready}},\\
 \Delta T_{\mathrm{SPLASH}}&=T_{\mathrm{SPLASH}}-T_{\mathrm{ready}}.
\end{aligned}
\label{eq:measured-switch-time}
\end{equation}
\autoref{tab:hotswitch-live} 
reports the measured B200 execution-time
p50 values and p50(p95) overheads at $B=16$ and 256K context.
Overhead quantiles are 
taken from the measured columns. 

\section{Limitations}
\label{app:limitations}

\paragraph{Graph capture.}
CUDA graphs bind the addresses of weights and KV buffers, so every supported
layout needs its own captured graphs, which \textsc{Splash} captures once at startup.
Capture time and resident graph memory must be accounted for at deployment.

\paragraph{Memory headroom.}
A switch holds one layer's buffers beyond the larger of the source and
destination footprints (\autoref{eq:transition-memory}). Near full
memory, \textsc{Splash} defers the switch or holds admission until that headroom
exists.

\paragraph{Imperfect overlap.}
Transfers share interconnect and memory bandwidth with the switching step's
own collectives and computation, which can slow that step. A switch that must
move long histories, such as restoring replicated MLA KV when entering TP,
can carry more bytes per layer than one layer's computation hides; the
excess lengthens the switching step (\autoref{eq:switch-cost}), and the
scheduler charges it to the switch cost.

\paragraph{Short-lived regimes.}
A switch pays off only if the destination's advantage outlasts the switch.
When the workload oscillates faster than that, the switching margin keeps
the current layout and the opportunity goes unused.

\paragraph{Implementation scope.}
The DOP transfer path requires equal TP and attention-DP degrees with one
rank per owner. Nesting a CP group inside a DOP owner, speculative decoding,
and heterogeneous workers are extensions beyond this design.

\section{DCU Evaluation}
\label{app:dcu-evaluation}

\subsection{Experimental setup}
\label{app:dcu-setup}

The fixed-layout study uses GLM-5.3-Flash 300B INT8 weights
(11 MLA and 34 KDA layers) with FP8 KV in SGLang on BW1000 DCUs.
A DCU node runs prefill; sixteen DCUs across two nodes run
DP-attention decode. \autoref{app:repro} gives the deployment settings.

\subsection{Fixed-layout performance}
\label{app:dcu-regimes}

Four DCU sweeps cover eight input lengths from 1K to
512K at $B=16,32$ and nine batch sizes
$B\in\{1,4,8,16,32,64,128,256,512\}$ at each of 32K and 64K input,
with $B$ requests at client concurrency $B$. All four layouts run at every
position, giving 136 measurements. Every request generates 1,024 output tokens, and K denotes
1,024 tokens. Total throughput is
$B(L_{\mathrm{in}}+1{,}024)/T_{\mathrm{batch}}$ over the batch end-to-end
time, which includes queueing, prefill, KV transfer, and decode; 
each plotted point averages 10 runs.
The 512K cases stress performance beyond the nominal 256K context;
\autoref{tab:dcu-full-sweep} lists all 136 values.

\begin{figure}[htbp]
\centering
\includegraphics[width=\linewidth]{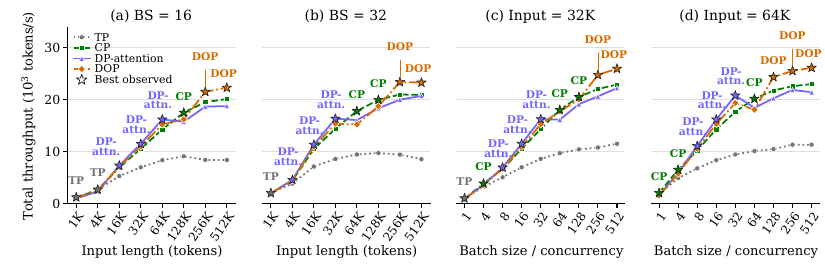}
\caption{%
Each layout leads a regime; DOP leads at 256K--512K and at the largest
batches.
GLM-5.3-Flash with common DP-attention decode and 1K output
tokens per request: (a, b) input-length sweeps at $B=16,32$; (c, d)
batch-size sweeps at 32K and 64K input; sampled positions are equally spaced.
Stars and labels mark the best recorded layout at each position.}
\label{fig:dcu-appendix-regime-crossover}
\label{fig:dcu-appendix-regime-batch}
\end{figure}

At 256K--512K, DOP reaches 21,539--23,436 tokens/s,
10.1--11.5\% above CP, the strongest alternative. It also leads at
$B=256,512$ for 32K input and $B=128,256,512$ for 64K input,
exceeding CP by 11.8--13.6\% at these five positions.
At $B=8,16,32$, however, DOP trails DP-attention by 3.1--6.9\%
in both batch sweeps despite its larger KV pool.

The independently recorded sweeps differ at 64K/$B=32$: CP leads in
\autoref{fig:dcu-appendix-regime-crossover}b, and DP-attention in (d).

\subsection{KV capacity and request admission}
\label{app:dcu-capacity}

Because it shards the projections, DOP holds a 19.0\% larger KV pool per
DCU than DP-attention with GLM-5.3-Flash (1,118,656 versus 940,224
tokens per owner; \autoref{tab:capacity}). The table compares effective
capacity across the device group in distinct tokens, accounting for KV replication.

This capacity raises throughput once it changes admission.
In a co-located DCU run, which interleaves chunked prefill with
decode for 240,000-token inputs and 512-token outputs, DOP overtakes DP-attention between 24 requests (17,526.79 versus
19,084.86 input tokens/s) and 32 (19,265.96 versus 16,347.98), because at
32 requests DOP admits four per owner while DP-attention admits three and
queues one.

\subsection{Switch primitive costs}
\label{app:dcu-switch-primitives}

We report single-layer directional measurements for GLM-5.3-Flash on
BW1000 DCUs, with four nonzero weight costs and seven nonzero KV costs.

\begin{figure}[t]
\centering
\includegraphics[width=\linewidth]{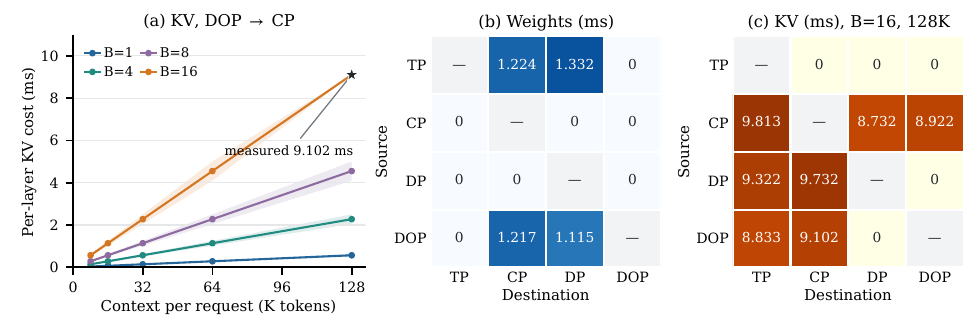}
\caption{Per-layer network costs.
(a) KV cost of DOP$\to$CP versus context, with linear $Bs$ scaling
summarizing the trend. The star marks the $B=16$, 128K reference;
bands show approximately $\pm10\%$ variation across 20 runs per case.
(b) Weight cost of the twelve switches; rows are sources, columns
destinations, and 0 marks no network transfer.
(c) KV cost at $B=16$ and 128K. DP denotes DP-attention.
All measurements use GLM-5.3-Flash on BW1000 DCUs.}
\label{fig:dcu-appendix-hotswitch-preparation}
\label{fig:dcu-appendix-hotswitch-operators}
\end{figure}

Every direction uses a discard, all-gather, or all-to-all per state and
layer (\autoref{fig:splash}, right). A discard has zero network cost;
\autoref{fig:dcu-appendix-hotswitch-preparation} reports the two collectives.

\paragraph{Weights cost a fixed amount.}
The four weight all-gathers, from TP or DOP into DP-attention or CP,
each move $(1-1/T)$ of one layer's projections per rank and take
1.115--1.332\,ms per layer
(\autoref{fig:dcu-appendix-hotswitch-preparation}b).
Their payload is independent of batch size and retained context.

\paragraph{KV cost scales with context.}
The seven KV all-gathers and all-to-alls take 8.732--9.813\,ms per layer
at $B=16$ and 128K tokens per request
(\autoref{fig:dcu-appendix-hotswitch-preparation}c).
The linear payload model for DOP$\to$CP uses its 9.102\,ms measurement:
\[
 \widehat t_K(B,s)=9.102\,\frac{B}{16}\frac{s}{131{,}072}\ \mathrm{ms}.
\]
At $B=16$, this gives an estimated 1.138\,ms at 16K
(\autoref{fig:dcu-appendix-hotswitch-preparation}a).
Each case is measured 20 times; the bands show the observed run-to-run
variation of approximately $\pm10\%$.

\subsection{Live switching overhead}
\label{app:dcu-live-overhead}

\autoref{tab:dcu-appendix-hotswitch-live} gives the measured DCU summaries
for all twelve directed switches at $B=16$ and 128K context tokens per
request. Blocking, \textsc{Splash}, and target-ready use the definitions
in \autoref{sec:live-overhead}. 
Here $T$ denotes the measured execution interval of one step, and $\Delta T$
the measured overhead over target-ready.

\begin{table}[t]
\caption{Supplied DCU switching summaries ($B=16$, 128K context
tokens per request). Times are milliseconds: $T$ is the timed interval's p50;
$\Delta T$ is p50(p95) overhead over target-ready (Ready). Block.\ is
blocking and DP denotes DP-attention.}
\label{tab:dcu-appendix-hotswitch-live}
\centering\fontsize{7.2}{9}\selectfont
\setlength{\tabcolsep}{1.1pt}
\begin{adjustbox}{max width=\linewidth}
\begin{tabular}{@{}lrrrrr@{\hspace{6pt}}lrrrrr@{}}
\toprule
 & \multicolumn{3}{c}{$T$, p50} & \multicolumn{2}{c}{$\Delta T$, p50(p95)}
 & & \multicolumn{3}{c}{$T$, p50} & \multicolumn{2}{c}{$\Delta T$, p50(p95)} \\
\cmidrule(lr){2-4}\cmidrule(lr){5-6}\cmidrule(lr){8-10}\cmidrule(lr){11-12}
Direction & Ready & Block. & \textsc{Splash} & Block. & \textsc{Splash}
 & Direction & Ready & Block. & \textsc{Splash} & Block. & \textsc{Splash} \\
\midrule
$\mathrm{TP}\!\to\!\mathrm{CP}$ & 694.85 & 790.76 & 696.53 & 95.91(176.16) & 1.68(3.30) & $\mathrm{DP}\!\to\!\mathrm{TP}$ & 1612.02 & 1901.83 & 1623.03 & 289.81(511.96) & 11.01(19.59) \\
$\mathrm{TP}\!\to\!\mathrm{DP}$ & 720.40 & 813.39 & 722.09 & 92.99(168.47) & 1.69(3.35) & $\mathrm{DP}\!\to\!\mathrm{CP}$ & 699.10 & 987.05 & 709.87 & 287.95(498.17) & 10.77(19.75) \\
$\mathrm{TP}\!\to\!\mathrm{DOP}$ & 611.54 & 611.57 & 611.56 & 0.03(0.06) & 0.02(0.04) & $\mathrm{DP}\!\to\!\mathrm{DOP}$ & 621.47 & 621.49 & 621.49 & 0.02(0.04) & 0.02(0.04) \\
$\mathrm{CP}\!\to\!\mathrm{TP}$ & 1543.94 & 1834.96 & 1555.04 & 291.02(516.72) & 11.10(20.26) & $\mathrm{DOP}\!\to\!\mathrm{TP}$ & 1534.59 & 1820.64 & 1545.20 & 286.05(504.06) & 10.61(18.80) \\
$\mathrm{CP}\!\to\!\mathrm{DP}$ & 715.27 & 994.71 & 725.95 & 279.45(517.48) & 10.68(18.78) & $\mathrm{DOP}\!\to\!\mathrm{CP}$ & 692.91 & 1075.84 & 704.67 & 382.93(659.08) & 11.76(20.11) \\
$\mathrm{CP}\!\to\!\mathrm{DOP}$ & 620.30 & 910.63 & 631.42 & 290.33(517.24) & 11.12(19.04) & $\mathrm{DOP}\!\to\!\mathrm{DP}$ & 723.57 & 818.80 & 725.29 & 95.23(176.48) & 1.72(3.25) \\
\bottomrule
\end{tabular}
\end{adjustbox}
\end{table}

Across all twelve directions, \textsc{Splash} adds 0.02--11.76\,ms
at p50 and 0.04--20.26\,ms at p95; its median overhead stays
below 1.8\% of the corresponding target-ready interval. For the ten directions
requiring network transfers, median overhead falls by 96.17--98.25\%
relative to blocking.

The aggregate quantiles do not separate initial preparation,
later-layer readiness waits, and computation interference.

\clearpage
\section{H200 Evaluation}
\label{app:h200-evaluation}

\subsection{Experimental setup}
\label{app:h200-setup}

We compare TP, CP, DP-attention, and DOP using DeepSeek-V3.2 on H200
with one prefill instance and one decode instance (1P1D). The fixed-layout
study varies the prefill layout. Every request generates 1,024 output
tokens, and K denotes 1,024 tokens. The switching studies use $B=16$
and 128K context tokens per request.

\subsection{Fixed-layout performance}
\label{app:h200-regimes}

Four sweeps cover seven input lengths from 1K to 256K at $B=16,32$,
eight batch sizes from 1 to 256 at 32K input, and seven batch sizes from
1 to 128 at 64K input. All four layouts run at each sampled position,
giving 116 measurements. Total throughput is
$B(L_{\mathrm{in}}+1{,}024)/T_{\mathrm{batch}}$ over the batch end-to-end
time. 
Each point averages 10 runs;
\autoref{tab:h200-full-sweep} lists all values.

Both input-length sweeps favor TP at 1K, DP-attention at 4K--16K,
and CP at 32K--64K. At $B=16$, CP also leads at 128K, while DOP
leads at 256K; at $B=32$, DOP leads at 128K--256K.
Across these three long-input points, DOP reaches 14,349--18,584 tokens/s,
4.2--10.3\% above CP, the strongest alternative.
In the batch sweeps, DOP first leads at $B=128$ for 32K input and
$B=64$ for 64K input. Its advantages at the two largest sampled batches
are 5.8--6.1\% and 10.4--12.1\%, respectively.
DOP leads at seven of the 29 sampled positions, while the other layouts
remain faster elsewhere.

Overlapping configurations retain their independently recorded results.
At 32K/$B=16$, CP leads in the input-length sweep, whereas DP-attention
leads in the batch-size sweep. 
These fixed-layout comparisons motivate workload-aware selection.

\subsection{Switch primitive costs}
\label{app:h200-switch-primitives}

The switching summaries compare all twelve directed transitions at
$B=16$ and 128K context tokens per request. 

\begin{figure}[t]
\centering
\includegraphics[width=\linewidth]{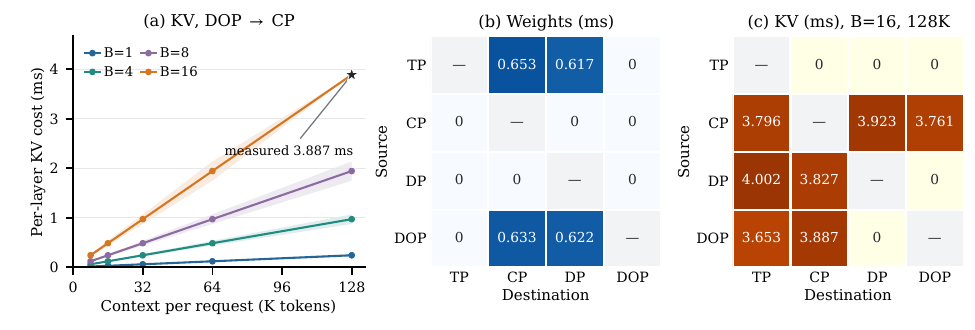}
\caption{Supplied H200 per-layer network costs.
(a) DOP$\to$CP KV cost, with linear $Bs$ scaling summarizing the trend.
The star marks the $B=16$, 128K reference; bands show approximately
$\pm10\%$ variation across 20 runs per case.
(b) Weight costs of the twelve switches.
(c) KV costs at $B=16$ and 128K context per request.
Rows are sources, columns are destinations, and 0 denotes no network
transfer. DP denotes DP-attention.}
\label{fig:h200-appendix-hotswitch-preparation}
\label{fig:h200-appendix-hotswitch-operators}
\end{figure}

\autoref{fig:h200-appendix-hotswitch-preparation} reports four nonzero
weight all-gather costs of 0.6174--0.6526\,ms per layer and seven nonzero
KV collective costs of 3.6528--4.0020\,ms per layer.
Linear scaling in $Bs$ from the DOP$\to$CP reference of 3.8870\,ms
summarizes the context trend. Each case is measured 20 times, with
observed run-to-run variation of approximately $\pm10\%$.

\subsection{Live switching overhead}
\label{app:h200-live-overhead}

\autoref{tab:h200-appendix-hotswitch-live} compares all twelve directed
switches. Blocking, \textsc{Splash}, and target-ready use the definitions
in \autoref{sec:live-overhead}. 
$T$ denotes the measured execution interval of one step, and $\Delta T$ the
measured overhead over target-ready.

\begin{table}[t]
\caption{Supplied H200 switching summaries at $B=16$ and
128K context tokens per request. Units are milliseconds.
$T$ is the timed interval's p50; $\Delta T$ reports p50(p95) overhead over
the matched target-ready reference. Block.\ denotes the blocking
baseline and DP denotes DP-attention.}
\label{tab:h200-appendix-hotswitch-live}
\centering\fontsize{7.2}{9}\selectfont
\setlength{\tabcolsep}{1.1pt}
\begin{adjustbox}{max width=\linewidth}
\begin{tabular}{@{}lrrrrr@{\hspace{6pt}}lrrrrr@{}}
\toprule
 & \multicolumn{3}{c}{$T$, p50} & \multicolumn{2}{c}{$\Delta T$, p50(p95)}
 & & \multicolumn{3}{c}{$T$, p50} & \multicolumn{2}{c}{$\Delta T$, p50(p95)} \\
\cmidrule(lr){2-4}\cmidrule(lr){5-6}\cmidrule(lr){8-10}\cmidrule(lr){11-12}
Direction & Ready & Block. & \textsc{Splash} & Block. & \textsc{Splash}
 & Direction & Ready & Block. & \textsc{Splash} & Block. & \textsc{Splash} \\
\midrule
$\mathrm{TP}\!\to\!\mathrm{CP}$ & 692.85 & 781.58 & 694.31 & 88.73(185.40) & 1.46(2.67) & $\mathrm{DP}\!\to\!\mathrm{TP}$ & 1408.22 & 1929.61 & 1415.84 & 521.39(876.42) & 7.62(15.05) \\
$\mathrm{TP}\!\to\!\mathrm{DP}$ & 765.24 & 858.42 & 766.75 & 93.18(177.53) & 1.51(2.79) & $\mathrm{DP}\!\to\!\mathrm{CP}$ & 708.92 & 1217.19 & 717.32 & 508.26(846.18) & 8.39(14.96) \\
$\mathrm{TP}\!\to\!\mathrm{DOP}$ & 646.78 & 646.81 & 646.80 & 0.03(0.07) & 0.02(0.07) & $\mathrm{DP}\!\to\!\mathrm{DOP}$ & 652.48 & 652.50 & 652.50 & 0.02(0.10) & 0.02(0.08) \\
$\mathrm{CP}\!\to\!\mathrm{TP}$ & 1462.84 & 1975.32 & 1470.68 & 512.48(851.61) & 7.84(15.31) & $\mathrm{DOP}\!\to\!\mathrm{TP}$ & 1505.49 & 2022.33 & 1513.46 & 516.84(831.71) & 7.97(14.52) \\
$\mathrm{CP}\!\to\!\mathrm{DP}$ & 727.91 & 1231.08 & 735.96 & 503.18(883.93) & 8.06(14.13) & $\mathrm{DOP}\!\to\!\mathrm{CP}$ & 697.16 & 1314.54 & 706.81 & 617.38(1122.60) & 9.65(17.07) \\
$\mathrm{CP}\!\to\!\mathrm{DOP}$ & 670.15 & 1165.87 & 678.43 & 495.72(817.95) & 8.27(14.77) & $\mathrm{DOP}\!\to\!\mathrm{DP}$ & 779.45 & 870.01 & 780.88 & 90.56(178.69) & 1.43(2.71) \\
\bottomrule
\end{tabular}
\end{adjustbox}
\end{table}

\textsc{Splash}'s additional time is
0.02--9.65\,ms at p50 and 0.07--17.07\,ms at p95;
its p50 overhead is below 1.39\% of the corresponding ready p50.
For the ten directions requiring network transfers, it reduces p50
overhead by 98.33--98.54\% relative to blocking.
The aggregate quantiles do not separate first-layer preparation,
later readiness waits, and computation interference.

\end{document}